\documentclass[a4paper,fleqn,usenatbib]{mnras}

\usepackage{newtxtext,newtxmath}

\usepackage[T1]{fontenc}
\usepackage{ae,aecompl}

\usepackage{graphicx}	
\usepackage{amsmath}	
\usepackage{amssymb}	
\usepackage{adjustbox}
\usepackage{multirow}
\usepackage{soul}
\usepackage{subcaption}
\title[Convex X-ray Spectra of PKS\,2155-304]{Convex X-ray Spectra of PKS\,2155-304 and Constraints on the Minimum Electron Energy}

\author[J. Sitha et al.]{Sitha K. Jagan$^{1}$\thanks{E-mail: sithajagan@gmail.com},
S. Sahayanathan$^{2}$\thanks{E-mail: sunder@barc.gov.in},
Frank M. Rieger$^{3,4}$\thanks{E-mail: frank.rieger@mpi-hd.mpg.de}
and C. D. Ravikumar$^{1}$
\\
$^{1}$Department of Physics, University of Calicut, Malappuram-673635, India\\
$^{2}$Astrophysical Sciences Division, Bhabha Atomic Research Centre, Mumbai - 400085, India\\
$^{3}$ZAH, Institute of Theoretical Astrophysics, University of Heidelberg, Philosophenweg 12, 69120 Heidelberg, Germany\\
$^{4}$Max-Planck-Institut f$\ddot{u}$r Kernphysik, P.O. Box 103980, 69029 Heidelberg, Germany
}

\date{Accepted XXX. Received YYY; in original form ZZZ}

\pubyear{2021}

\begin{document}
\label{firstpage}
\pagerange{\pageref{firstpage}--\pageref{lastpage}}
\maketitle

\begin{abstract}
The convex (concave upward) high-energy X-ray spectra of the blazar PKS\,2155-304, observed by \emph{XMM-Newton}, is interpreted as the signature of sub-dominant inverse Compton emission. The spectra can be well fitted by a superposition of two power-law contributions which imitate the emission due to synchrotron and inverse Compton processes. The methodology adopted enables us to constrain the photon energy down to a level where inverse Compton emission begins to contribute. We show that this information supplemented with knowledge of the jet Doppler factor and magnetic field strength can be used to constrain the low-energy cutoff $\gamma_{\rm min}m_{\rm e} c^2$ of the radiating electron distribution and the kinetic power $P_{\rm j}$ of the jet. We deduce these quantities through a statistical fitting of the broadband spectral energy distribution of PKS\,2155-304 assuming synchrotron and synchrotron self Compton emission mechanisms. Our results favour a minimum Lorentz factor for the non-thermal electron distribution of $\gamma_{\rm min} \gtrsim 60$, with a preference for a value around $\gamma_{\rm min} \simeq 330$. The required kinetic jet power is of the order of $P_{\rm j} \sim 3\times 10^{45}$ erg s$^{-1}$ in case of a heavy, electron-proton dominated jet, and could be up to an order of magnitude less in case of a light, electron-positron dominated jet. When put in context, our best fit parameters support the X-ray emitting part of blazar jets to be dominated by an electron-proton rather than an electron-positron composition.
\end{abstract}




\section{Introduction}

A prominent morphological feature of the radio-loud type AGN is the presence of relativistic jets extending up to megaparsec scales. 
Blazars are the class of radio-loud AGN in which the jet is aligned close to the line of sight of the 
observer. Hence the emission from blazars is significantly Doppler-boosted. Their spectrum is 
predominantly non-thermal in nature, extending from radio to GeV and TeV energies 
\citep[e.g.,][]{2010ApJ...716...30A,2013APh....43..103R}. Blazars are further classified into flat spectrum 
radio quasars (FSRQs) or BL Lac type objects (BL Lacs) depending upon the presence or absence of 
emission line features.
The spectral energy distribution (SED) of blazars is typically characterized by a double-hump structure 
with the low-energy component peaking around IR/optical/UV energies and the high-energy component peaking 
around gamma-ray energies \citep[e.g.,][]{1998MNRAS.299..433F,2017MNRAS.469..255G}. 
The low-energy emission is well understood to be synchrotron emission from a relativistic non-thermal 
electron distribution; whereas, the high-energy emission is often interpreted as due to inverse Compton 
(IC) scattering of low-energy photons by the same electron distribution \citep[e.g.,][]{1992ApJ...397L...5M,
1985ApJ...298..114M,1987ApJ...322..650B,1993ApJ...416..458D}. 
The peak energy of the synchrotron spectral component is used to sub-divide BL Lacs into low energy 
peaked BL Lacs (LBLs: peaking at IR/optical), intermediate energy peaked BL Lacs (IBLs: peaking at 
optical/UV) and high energy peaked BL Lacs (HBLs: peaking at UV/soft X-ray) \citep{1995ApJ...444..567P,
1998MNRAS.299..433F}. Within the leptonic framework, the gamma-ray spectra of HBLs are well reproduced by synchrotron 
self-Compton (SSC) emission. The target photons for the IC scattering in this case are the synchrotron photons themselves. 

Spectral modelling of HBLs by synchrotron and SSC processes suggests that the underlying electron distribution 
may be closely related to a power-law/broken power-law type function resulting from Fermi acceleration and 
radiative cooling processes \citep[e.g.,][]{1998A&A...333..452K,1999APh....11...45K}.
However, high resolution X-ray spectra can exhibit significant deviations from a power-law and are often 
better reproduced by a log-parabola function \citep{2008A&A...478..395M}. Such curved spectra demand 
a modified electron distribution which can be achieved by assuming energy-dependent acceleration or 
particle diffusion time scales \citep{2011ApJ...739...66T,2007A&A...466..521T,2018MNRAS.478L.105J,
2020MNRAS.499.2094G}. 


Even after substantial progress in understanding the emission processes in blazars, the matter content 
and kinetic power of blazar jets are still open questions \citep[e.g.,][]{2008MNRAS.385..283C,
2016ApJ...831..142M,2014Natur.515..376G,2015MNRAS.450L..21Z}. Though the detection of neutrinos 
provides indications for the presence of hadrons in blazar jets \citep{2018Sci...361.1378I,2014ApJ...793L..18T}, 
their signature in the emission spectra is largely unknown since models advocating a hadronic origin of 
high-energy spectra pose serious energy constraints \citep{2015MNRAS.450L..21Z}.
Leptonic models are largely successful in explaining the broadband SED of blazars especially during
the flare. 
These models generally assume the protons to be cold and contribute only to the jet kinetic power 
\citep{2008MNRAS.385..283C}. However, if their number is equal to that of the non-thermal electrons 
(heavy jet), the predicted jet power could exceed the accretion disk luminosity \citep{2014Natur.515..376G}.

For a power-law or power-law type electron distribution $N(\gamma)$ of index $p>1$, the total electron number 
density $N_{\rm tot}$ is determined by the differential number density at the minimum cut-off energy 
$\gamma_{\rm min}$, i.e., $N_{\rm tot} \simeq N(\gamma_{\rm min})~\gamma_{\rm min}/(p-1)$, $\gamma$ being 
the electron Lorentz factor. Under a Fermi-type acceleration process, this energy can be 
related to the energy of the electron population injected into the acceleration region 
\citep{1998A&A...333..452K,2018ApJ...861..152K}. As mentioned earlier, the knowledge of the total electron number density 
can also provide an estimate of the blazar jet mass density by assuming an appropriate fraction of hadrons. 
This, along with broadband spectral modelling, can provide clues on the kinetic power of blazar jets. 
However, the estimation of $\gamma_{\rm min}$ based on spectral information is complicated by the fact that 
the related synchrotron emission usually falls into the radio regime which is significantly self-absorbed 
or contaminated by extended jet emission. 
On the other hand, the low-energy IC contribution may be overwhelmed by the dominant synchrotron 
emission. Still, an upper limit on the related minimum IC photon energy might be estimated from the 
transition energy in the broadband spectra, where the dominant contribution shifts from synchrotron to 
IC emission \citep{2016ApJ...827...55K}.

PKS\,2155-304 is a HBL located at a redshift of $z=0.116$ and well observed from radio to very high 
energy (VHE) gamma rays. Its synchrotron spectrum peaks at UV and the IC spectral component 
peaks at GeV energies \citep{2008ApJ...682..789Z,2016ApJ...831..142M}. Its broadband spectra can 
be reproduced reasonably well by synchrotron and SSC emission of a broken power-law type electron 
distribution \citep{2016ApJ...831..142M,2020A&A...639A..42A}. The source is known to exhibit rapid 
variability with doubling time as short as a few minutes, suggesting that the emission region is located 
very close to the supermassive black hole powering the AGN \citep{2007ApJ...664L..71A,
2010A&A...520A..23R}.

The mild curvature observed in the high resolution X-ray spectra can be interpreted as an outcome 
of the energy-dependence of the particle escape time-scale. However, during certain epochs, 
the X-ray spectral curvature reverses the sign indicating a convex (concave upward) spectrum suggesting 
the spectral contribution of a sub-dominant IC component in addition to the synchrotron component \citep{2018MNRAS.478L.105J}.

The contribution of IC emission to the X-ray spectra of PKS\,2155-304 can be inferred from 
\emph{XMM-Newton} observation, where the spectrum has been found to harden beyond 4
 keV. This was first shown by \cite{2008ApJ...682..789Z}, who used a broken power-law spectral 
 fit to highlight the presence of an IC component. 
\emph{NuSTAR} observations of PKS\,2155-304 at 0.5-60 keV also reveal that the IC spectral 
component dominates over the synchrotron spectrum beyond $\approx$ 6 keV 
\citep{2016ApJ...831..142M}. This has been demonstrated by fitting the spectrum with a broken
power-law, a combination of two power-law functions (double power-law) and log parabola--power-law 
combination, respectively.
\emph{NuSTAR} observation of yet another HBL, MKN\,421 also showed a convex X-ray spectrum 
during a low-flux state that can be well reproduced by a double power-law function 
\citep{2016ApJ...827...55K}.
In general, the presence of convex X-ray spectra has also been witnessed for many IBLs and 
LBLs and are interpreted as the signature of the IC spectral component \citep[e.g.,][]{2000A&A...354..431T,
2003ApJ...584..153T,2016MNRAS.458...56W,2018MNRAS.473.3638G}.

Alternatively, a convex X-ray spectrum may also be an outcome of a high energy pile-up of the 
emitting electron distribution. Such a high energy excess in the electron distribution can occur 
in at least two scenarios. When the confinement time of the electrons in the region of particle 
acceleration is much longer than the radiative loss timescales, the electrons will eventually get 
accumulated at the maximum energy giving rise to a relativistic Maxwellian tail at high energies 
\citep[e.g.,][]{1985A&A...143..431S,2000MNRAS.312..579O,2008ApJ...681.1725S,2011ApJ...740...64L}. 
Another scenario could be, when the electron energy loss rate at high energy is dominated 
by IC scattering of the external photon field happening in the Klein-Nishina regime 
\citep{2002ApJ...568L..81D,2005MNRAS.363..954M}.\

While in the case of PKS\,2155-304 the high energy emission is thought to be dominated by 
SSC processes rather than external Compton, it is often not straightforward to differentiate the 
process responsible for the convex X-ray spectra.
Nevertheless, the contribution of a sub-dominant Compton spectral component can be identified 
through a statistical study between the X-ray spectral curvature and the simultaneous low energy 
gamma-ray spectral index of the Compton spectrum.

In the present work, we perform a detailed study of the convex X-ray spectra of PKS\,2155-304 
observed by \emph{XMM-Newton}. The convexness observed in the 0.6-10 keV X-ray spectrum 
is interpreted as the signature of a sub-dominant SSC contribution. We extend this along with 
broadband spectral modelling to constrain the low-energy cut-off of the emitting electron distribution 
and the source energetics. 

The paper is organised as follows: In \S~\ref{Sec2}, we explain the observations and data reduction 
procedure. Epochs during which the X-ray spectrum is convex are identified in \S~\ref{Sec3}.
In \S~\ref{sec:xcom} and \ref{sec:jeteng}, we show that this can be interpreted as the result of a 
combination of synchrotron and SSC spectral components using a double power-law function. 
In \S~\ref{sec:sedfit} we describe the broadband spectral fitting of the source and the estimation of 
jet power. The outcome is discussed in \S~\ref{sec:discuss}.
Throughout this work, we adopt a cosmology with $\Omega_{\rm m} = 0.3$, $\Omega_{\rm \Lambda} = 
0.7$ and $H_{\rm 0} = 70$ km s$^{-1}$ Mpc$^{-1}$.

\section{OBSERVATIONS AND DATA REDUCTION}\label{Sec2}

PKS\,2155-304 has been observed extensively by \emph{XMM-Newton}, with 37 \emph{XMM-Newton} 
master observations available in the \emph{HEASARC} archive from 2000-2014. 
For our X-ray study, we have taken the \emph{XMM-Newton}'s European Photon Imaging Camera (EPIC)-PN 
data only and avoided EPIC-MOS data. The reason for this being less sensitivity and quantum efficiency, and 
the chance of pile up in EPIC-MOS data during the observation of bright sources. Among 37 \emph{XMM-Newton} 
observations, only 22 PN observations were available during which the camera was operated in the small window 
imaging mode. The \emph{XMM-Newton} Science Analysis System (SAS version 14.0) with the latest calibration 
files was used for the data reduction.

The standard procedures were followed to reprocess the Observation Data File (ODF). The \emph{XMM-Newton} 
SAS pipeline command \emph{epchain} is used to produce the calibrated photon event files for the PN camera. 
In the selected 0.2-10.0 keV energy range, we considered both single and double events (PATTERN$\le4$) of PN 
data which were flagged good quality (FLAG=0) for processing. We extracted the light curves in the high energy 
range 10-12 keV in order to check the background particle flaring. An appropriate threshold rate was chosen 
from the light curve to omit the background flaring period and a ``good time interval (GTI)" event list was created. 
A circular region of size $40^{\prime\prime}$ around the source was selected for extracting the source spectrum and 
two circular regions of the same size were chosen as the background. These background regions were selected 
from the same CCD chip of source region such that the source photon contribution was negligible within these regions.

The chance of pile up is more probable for bright sources like PKS\,2155-304 and hence, we used the SAS task 
\emph{epatplot} to examine the pile up effect in all observations. We found that certain observations are affected 
by the pile up. In order to avoid these pile up effects, we considered an annulus region around the bright source 
with an inner radius of $10^{\prime\prime}$ and outer radius varying from $38^{\prime\prime}$ to $40^{\prime\prime}$. 
The range of the outer radius was chosen such that the source selection region remained within the frame of the 
CCD chip. Only the annulus region was taken into account for source count extraction and the inner region was excluded. 

The SAS tool \emph{specgroup} was used for grouping the spectral channels. We imposed a condition of 
having minimum 100 counts in each group. PKS\,2155-304 being a bright source, there can be large number of 
counts in the spectrum which in turn results in large number of bins. To avoid the chance of oversampling, we fixed 
the oversampling value to 5 so that the energy resolution cannot be covered by more than 5 grouped bins. The 
Galactic hydrogen column density ($N_{\rm {H}}=1.71 \times 10^{20} \ \rm cm^{-2}$) toward PKS\,2155-304 was kept 
constant for all spectral fits \citep{2014MNRAS.444.3647B}.
 
We also used the simultaneous optical observation of the source by Optical Monitor (OM) onboard \emph{XMM-Newton}. 
Reprocessing of the OM data was carried out by the SAS metatask \emph{omichain}. From the combined source list, our source of interest was obtained by cross matching its RA and Dec with the coordinates of all sources in the list. Unlike X-rays, optical/UV photons from the jet can be contaminated by the galactic photons at the same wavelength. This was treated as systematic error, and similar to \citet{2018MNRAS.478L.105J}, it was estimated by fitting the optical/UV data by a power-law function.  We found that adding 3 per cent systematic error in most of the data sets led to a better fit statistics. However, we had to reject 6 observations which either demanded higher systematic errors or lacked a minimum of 3 optical/UV flux points to perform the statistical fit. Finally, we got 16 observations which are used for the present study. Among these observations we selected only those observations which showed convex X-ray spectra (see \S~\ref{Sec3}). The details of these observations are given in Table \ref{tab:dblfit}. The galactic reddening correction was done for all the optical/UV data using the model \emph{UVRED} and fixing the parameter $ E_{\rm B-V}=0.019$ \citep{1979MNRAS.187P..73S, 2011ApJ...737..103S}.

\section{X-ray Spectral Curvature}\label{Sec3}
To identify the X-ray spectral curvature of PKS\,2155-304, the 22 epochs of \emph{XMM-Newton} PN observations 
over the energy range 0.6-10.0 keV were fitted with the X-ray Spectral Fitting Package (XSPEC) \citep{1996ASPC..101...17A} 
using a log-parabola function \citep{2004A&A...422..103M}.
This function is defined as 
\begin{align}\label{eq:logpar}
	F_{\rm lp}(\epsilon) = F_{\rm p} \left(\frac{\epsilon}{\epsilon_{\rm p}}\right)^{-\alpha-\beta \;{\rm log}(\epsilon/\epsilon_{\rm p})}\,,
\end{align} 
where $\epsilon$ is the photon energy (keV), $F_{\rm p}$ is the flux (photons cm$^{-2}$ s$^{-1}$ keV$^{-1}$) at photon energy $\epsilon_{\rm p}$ and the spectral shape is determined by the two parameters $\alpha$ and $\beta$. The parameter $\alpha$ governs the spectral slope
at $\epsilon_{\rm p}$ and $\beta$ is the parameter defining the spectral curvature. A negative value of $\beta$ 
indicates a convex spectrum that could hint at the presence of Compton spectral component or a high energy 
excess in the underlying electron distribution. In the 0.6-10.0 keV regime the spectrum of PKS\,2155-304 is 
dominated by the high energy end of the synchrotron emission, so that any IC contribution can result in mild 
negative value of $\beta$.

The log-parabola spectral fit to most of the observations resulted in positive $\beta$ suggesting a convex 
spectrum \citep{2018MNRAS.478L.105J}. 
Among the 22 observations, four (IDs 0158961401, 0411780101, 0411780201 and 0411780701) have
negative $\beta$ values indicative of a convex nature. The best fit parameters during these epochs are given in Table \ref{tab:logpar}. For three observations (0158961401, 0411780101 and 0411780201) the
negative values of $\beta$ are statistically significant. 

\begin{table*}
	\begin{center}
	\begin{tabular}{|c|c|c|c|c|c|}
	\hline
\multirow{2}{*}{Obs. ID} & Date of Observation  & \multicolumn{4}{c|}{logpar}\\ \cline{3-6} 
				& (yyyy.mm.dd) & $\alpha$ & $\beta$ & norm($10^{-3}$) & $\chi_{\rm red}^2$(dof) \\ \hline
0158961401 & 2006-05-01  & ${2.63\pm0.01}$ & ${-0.10\pm0.02}$ & ${14.97\pm0.06}$ &   1.06(173) \\
0411780101 & 2006-11-07  & ${2.58\pm0.01}$ & ${-0.10\pm0.02}$ & ${19.41\pm0.06}$  &   1.01(184) \\
0411780201 & 2007-04-22  & ${2.70\pm0.01}$ & ${-0.04\pm0.01}$ & ${46.32\pm0.10}$ &   1.11(248)  \\
0411780701 & 2012-04-28  & ${2.90\pm0.01}$ & ${-0.02\pm0.03}$ & ${5.63\pm0.02}$ &   1.11(149) \\
	\hline
	\end{tabular}
	\end{center}
	\caption{Log-parabola spectral fit results (photons cm$^{-2}$ s$^{-1}$ keV$^{-1}$) for the epochs with convex X-ray spectra (pivot energy fixed at 1 keV).}
	\label{tab:logpar}
\end{table*}


\section{Compton Spectral Signature in X-ray spectra}\label{sec:xcom}
To explore the contribution of a Compton component in the X-ray spectra of PKS\,2155-304, we assume that 
the synchrotron and SSC spectra at 0.6-10.0 keV can each be represented by a power-law. Accordingly, 
the cumulative spectrum ($\nu F_{\rm \nu}$) will be a double power-law defined by
\begin{align}\label{eq:doupow}
	F_{\rm dp}(\epsilon)=F_{\rm 0}\left[ \left(\frac{\epsilon}{\epsilon_{\rm m}}\right)^{-\Gamma_{\rm syn}} 
	+ \left(\frac{\epsilon}{\epsilon_{\rm m}}\right)^{\Gamma_{\rm com}}\right]\,,
\end{align}
where $\epsilon_{\rm m}$ is the photon energy at which the fluxes due to synchrotron and SSC processes are equal 
to $F_{\rm 0}$, and $-\Gamma_{\rm syn}$ and $\Gamma_{\rm com}$ are the synchrotron and SSC spectral indices,
respectively. 

Applying this, we find that the convex X-ray spectra of PKS\,2155-304, corresponding to the observation 
IDs 0158961401, 0411780101 and 0411780201 can be well-fitted with a double power-law function. 
However, a direct comparison between the fit statistics of the double power-law and the log-parabola functions
cannot be done due to the difference in the number of free parameters. 
The log-parabolic spectrum (eq.~{\ref{eq:logpar}}) is governed by three parameters ($F_{\rm p}$, $\alpha$ and $\beta$),
while the double power-law spectrum is governed by four parameters ($F_{\rm 0}$, $\epsilon_{\rm m}$, $\Gamma_{\rm syn}$ 
and $\Gamma_{\rm com}$). To progress, we reduce the number of free parameters of the double power-law 
function to three by fixing $\Gamma_{\rm com}$ at a value equal to the best fit spectral index of the simultaneous 
optical/UV spectrum. The rationale behind this being, if the 0.6-10.0 keV SSC spectrum is produced by the same 
power-law electron distribution responsible for the optical/UV emission also, then in the Thomson scattering regime 
both spectral indices will be equal. By this approach, the number of free parameters of the double power-law model 
is made equal to that of the log-parabola so that, a direct comparison of fit statistics is possible. 
The results of fit are given in Table \ref{tab:dblfit} and show that a double power-law model can also explain 
the convex X-ray spectrum as compared to a log-parabola model. 
The X-ray spectral curvature of PKS\, 2155-304 has also been studied by \citet{2017ApJ...850..209G}. Their
study also indicated the convex nature of the X-ray spectrum corresponding to the observation ID 0411780701. 
However, the F-test result suggests this curvature to be minimal. In additional to the $\chi^2$ test, we have 
performed F-test to obtain the statistical significance of log-parabola or double power-law. This analysis is performed 
for all the observations and the test results are given in Table \ref{tab:ftest}. We also find that the log-parabola 
spectral fit to the observation ID 0411780701 do not show any significant improvement over the simple power-law. 
Hence, we have excluded this observation from the rest of the study.

\begin{table*}
	\begin{center}
	\begin{tabular}{|c|c|c|c|c|c|c|c|}
	\hline
\multirow{2}{*}{Obs. ID} & \multicolumn{3}{c|}{PL} & \multicolumn{2}{c|}{logpar}  & \multicolumn{2}{c|}{double PL}\\ \cline{2-8} 
				& P & norm($10^{-3}$) & $\chi_{\rm red}^2$(dof) & F-test & P-value  & F-test & P-value\\ \hline
0158961401 & ${2.59\pm0.01}$ &  ${15.07\pm0.05}$ &   1.35(174)  & 49.22 & $4.95\times10^{-11}$ & 50.79 & $2.67\times10^{-11}$\\
0411780101 & ${2.55\pm0.01}$ &  ${19.58\pm0.06}$  &   1.39(185)  & 71.01 & $1.00\times10^{-14}$ & 67.29 & $3.95\times10^{-14}$ \\
0411780201 & ${2.69\pm0.00}$ & ${46.43\pm0.09}$ &   1.19(249)  & 19.04 & $1.88\times10^{-5}$ & 17.05 & $4.98\times10^{-5}$ \\
0411780701 & ${2.89\pm0.01}$ & ${5.63\pm0.02}$ &   1.11(150)   & 0.91 & 0.34 & 1.39 & 0.24\\
	\hline
	\end{tabular}
	\end{center}
	\caption{Power-law (PL) spectral fit and F-test results for the epochs with convex X-ray spectra. Columns 2 - 4 are the index, normalization 
	and reduced chi square (degrees of freedom) for PL fit. Columns 5 and 6 are the F statistic value and null-hypothesis probability for 
logparabola against PL model. Columns 7 and 8, are the F statistic value and null-hypothesis probability for 
double power-law against PL model.}
	\label{tab:ftest}
\end{table*}

\begin{table*}
	\begin{center}
	\begin{tabular}{c|c c c c|c c}
	\hline
Obs. ID &${\Gamma_{\rm syn}}$&log $\epsilon_{\rm v}$ (keV)&$F_{\rm 0.6 - 10 \ {keV}}$&$\chi_{\rm red}^2$(dof)&${\Gamma_{\rm com}}$($\chi_{\rm red}^2$)&$\kappa_{\rm lp}$\\ 
	\hline
0158961401  &  ${0.70\pm0.03}$ & ${1.93}^{+0.16}_{-0.14}$ & ${3.88\pm0.02}$  & 1.05(173) & ${0.17\pm0.06}$(1.77) & ${0.12\pm0.03}$ \\ 
0411780101  &  ${0.64\pm0.02}$ & ${1.67}^{+0.10}_{-0.08}$ & ${5.28\pm0.02}$  & 1.03(184) & ${0.26\pm0.06}$(1.81) & ${0.13\pm0.02}$ \\
0411780201  &  ${0.72\pm0.01}$ & ${2.08}^{+0.25}_{-0.12}$ & ${12.78\pm0.03}$ & 1.12(248) & ${0.32\pm0.05}$(1.98) & ${0.04\pm0.01}$ \\
$0411780701^\ast$ & ${0.89\pm0.01}$ &       -            & ${1.47\pm0.01}$  & 1.11(150) & ${0.28\pm0.05}$(1.22) & ${0.01}^{+ 0.02}_{-0.05}$ \\

\hline
      	\end{tabular}
	\end{center}
	\caption{Double power-law spectral fit results ($\nu F_{\rm \nu}$) for the epochs with convex X-ray spectra. $\epsilon_{\rm v}$ is the photon energy in keV corresponding to the valley in the SED (equation~\ref{eq:valley}). The index ${\Gamma_{\rm com}}$ is fixed to the value obtained from a power-law fit to the UV/optical spectra (see \S~\ref{Sec2}). $F_{\rm 0.6 - 10 \ {keV}}$ is the integrated flux over the energy range 0.6 to 10 keV ($10^{-11}$ erg $\rm cm^{-2}$ $\rm s^{-1}$). The reduced $\chi^{2}$ and degrees of freedom (dof) correspond to the best fit double power-law model. $\kappa_{\rm lp}$ is the curvature of the best fit log parabola model. The ${\Gamma_{\rm syn}}$ and $F_{\rm 0.6 - 10 \ {keV}}$ corresponding to observation $0411780701^\ast$ are obtained by a simple power-law fit.}
	\label{tab:dblfit}
\end{table*}

\begin{figure}
\includegraphics[scale=0.6,angle=270]{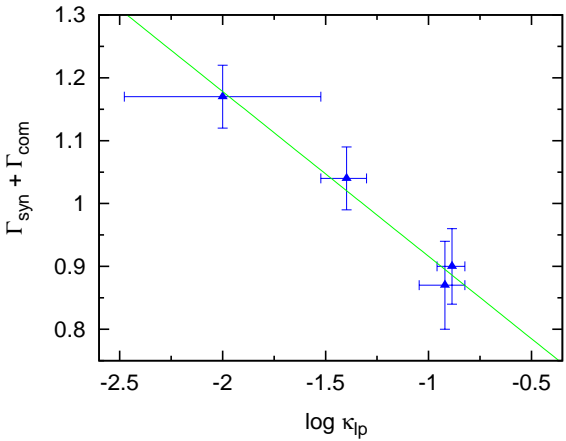}
\caption{Scatter plot between curvature of logparabola function ($\kappa_{\rm lp}$) with (${\Gamma_{\rm syn}} 
+ {\Gamma_{\rm com}}$) for the epochs with convex X-ray spectra. Solid green line is the best fit straight 
line to the data points (see text).}
\label{fig:indexflux}
\end{figure}

In principle, the spectral slope and the concavity of the double power-law function can be obtained from the
first and second derivatives of log\,$F_{\rm dp}$ with respect to log\,$\epsilon$, i.e., 
\begin{eqnarray}
{\rm log} F_{\rm dp}^{'} \equiv  
	\frac{d({\rm log}\,F_{\rm dp})}{d({\rm log}\,\epsilon)}  
	= \left(\frac{-\Gamma_{\rm syn}\,x_{\rm m}^{-\Gamma_{\rm syn}}
	   +\Gamma_{\rm com}\,x_{\rm m}^{\Gamma_{\rm com}}}{x_{\rm m}^{-\Gamma_{\rm syn}}+x_{\rm m}^{\Gamma_{\rm com}}}\right) \\
{\rm log} F_{\rm dp}^{''} \equiv \frac{d^2({\rm log}\,F_{\rm dp})}{d({\rm log}\,\epsilon)^2} 
        = \frac{(\Gamma_{\rm com}+\Gamma_{\rm syn})^2\,x_{\rm m}^{-\Gamma_{\rm syn}+\Gamma_{\rm com}}}{[x_{\rm m}^{-\Gamma_{\rm syn}}  
        +x_{\rm m}^{\Gamma_{\rm com}}]^2} >0\,,
\end{eqnarray}
where $x_{\rm m} =\epsilon/\epsilon_{\rm m}$. The positive value of the second derivative reaffirms the convex nature 
of the chosen double power-law function. The curvature $\kappa_{\rm dp}$ of the double power-law function 
is given by
\begin{align}\label{eq:dpcurv}
	\kappa_{\rm dp} =& \frac{{\rm log} F_{\rm dp}^{''}}{\left(1+{{\rm log} F_{\rm dp}^{'}}^2 \right)^{3/2}} \nonumber \\
	=& \frac{(\Gamma_{\rm com}+\Gamma_{\rm syn})^2 \left(x_{\rm m}^{-\Gamma_{\rm syn}}+x_{\rm m}^{\Gamma_{\rm com}}\right)
	x_{\rm m}^{-\Gamma_{\rm syn}+\Gamma_{\rm com}}}
	{\left[\left(x_{\rm m}^{-\Gamma_{\rm syn}}+ x_{\rm m}^{\Gamma_{\rm com}}\right)^2 + 
		\left(\Gamma_{\rm syn}\,x_{\rm m}^{-\Gamma_{\rm syn}}- \Gamma_{\rm com}\,x_{\rm m}^{\Gamma_{\rm com}}\right)^2\right]^{3/2}}\,.
\end{align}
The valley energy $\epsilon_{\rm v}$, corresponding to the minimum flux in the $\nu F_\nu$ representation can be 
obtained by setting equation (3) to zero, yielding
\begin{align}\label{eq:valley}
     \epsilon_{\rm v} = \epsilon_{\rm m}
                          \left(\frac{\Gamma_{\rm syn}}{\Gamma_{\rm com}}\right)^{\frac{1}{\Gamma_{\rm syn}+\Gamma_{\rm com}}}\,.
\end{align}
From equation~(\ref{eq:dpcurv}), $\kappa_{\rm dp}$ for the case $x_{\rm m} \ll 1$ will be 
\begin{align}\label{eq:dp_curv}
	\kappa_{\rm dp}(x_{\rm m}\ll1) \approx \frac{(\Gamma_{\rm syn}
	+\Gamma_{\rm com})^2}{{(1+\Gamma_{\rm syn}^2)^{3/2}}} x_{\rm m}^{\Gamma_{\rm syn}+\Gamma_{\rm com}}\,.
\end{align}
Hence in this case, $\kappa_{\rm dp}(x_{\rm m}\ll1)$ will decrease with increasing $\Gamma_{\rm syn}
+\Gamma_{\rm com}$. 
As mentioned before, the optical/UV spectral index could be used as an approximation for $\Gamma_{\rm com}$ while for $\Gamma_{\rm syn}$ we used the results from the double power-law fitting. For the observation 0411780701, $\Gamma_{\rm syn}$ was obtained through a simple power-law fit (Table \ref{tab:dblfit}). We calculated the X-ray spectral curvature at 1 keV using the log-parabola model instead of double power-law.  This is to avoid any possible bias as $\Gamma_{\rm syn}$ was also estimated assuming a double power-law. In case of a log-parabolic function, the curvature in $\nu F_{\rm \nu}$ representation can be obtained as
\begin{align}\label{eq:lpcurv}
	\kappa_{\rm lp} = 
	-\frac{2\beta}{\left\{1+\left[2-\alpha-2\beta\,{\rm log}\left(\frac{\epsilon}{\epsilon_{\rm p}}\right)\right]^2\right\}^{3/2}}\,.
\end{align}
We then study the dependence of $\kappa_{\rm lp}$ on the sum of optical/UV and X-ray spectral indices. This is done by redefining the 
log-parabola function in terms of $\kappa_{\rm lp}$ using equation~(\ref{eq:lpcurv}) and adding it as a local 
model in XSPEC. The best fit $\kappa_{\rm lp}$ values are given in column 7 of  Table \ref{tab:dblfit}. As can be seen, the spectral curvature can be as high as 0.15.
%
Equation~(\ref{eq:dp_curv}) suggests that log($\kappa_{\rm lp}$) varies linearly with $\Gamma_{\rm syn}+
\Gamma_{\rm com}$. In Figure \ref{fig:indexflux}, we plot these quantities and the least square fit, considering 
both the uncertainties \citep{1992nrfa.book.....P}, yielding $a=0.64\pm 0.14$ and $b = -0.28 \pm 0.12 $ with a goodness-of-fit (Q-value) of $0.89$\footnote {We fit a straight line of the form $y = a+bx$ to the set of points ($x_{\rm i}\pm\Delta x_{\rm i}$,$y_{\rm i}\pm\Delta y_{\rm i}$). If Q-value $>0.1$ the fit is trustable.}. 
Hence, this result supports our inference that the convex X-ray spectra of PKS\,2155-304 is an outcome of 
the superposition of a synchrotron and a SSC emission component. 

\begin{figure}
\includegraphics[scale=0.6,angle=270]{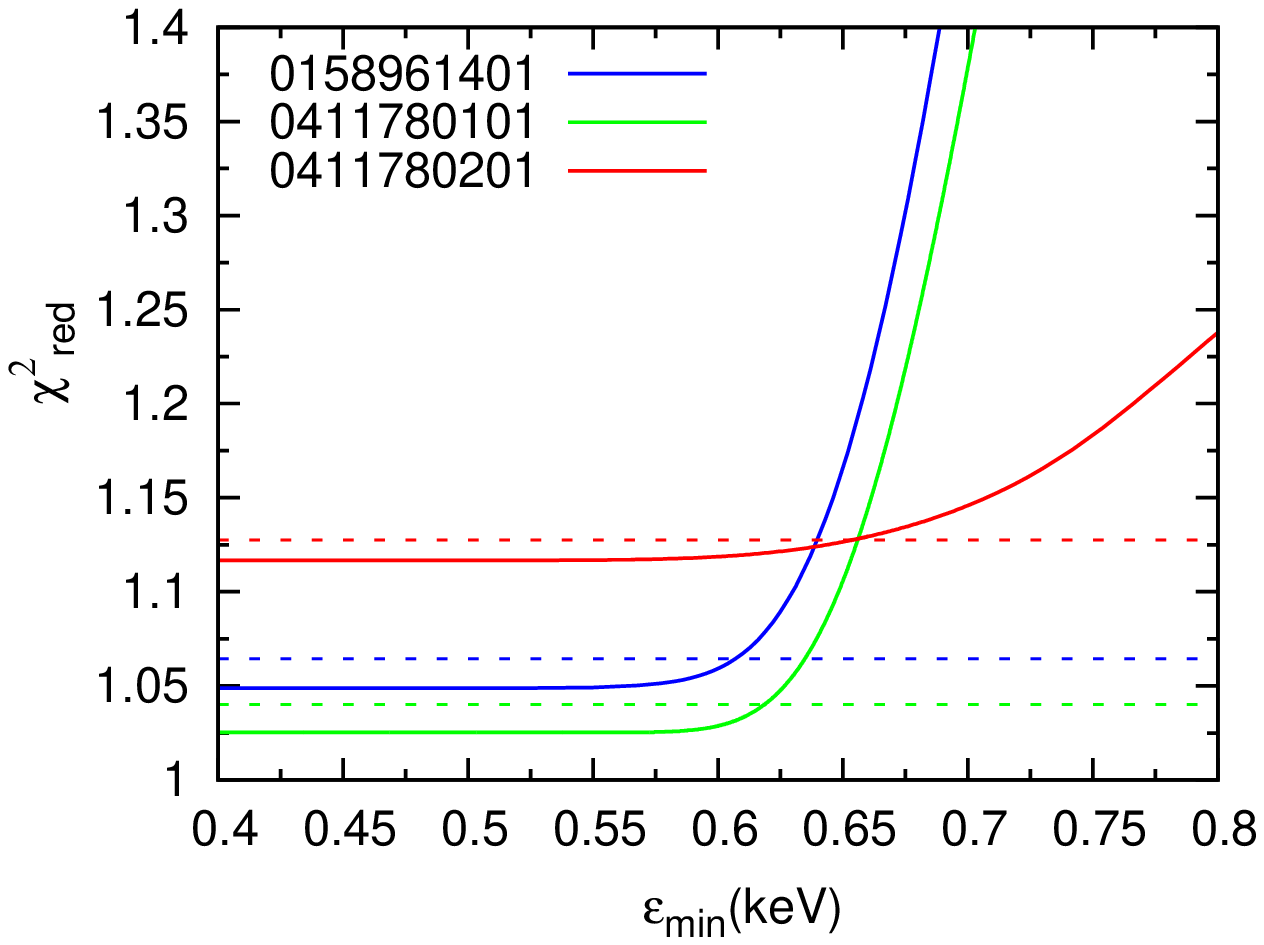}
\caption{Variation of $\chi^{2}_{\rm red}$ with $\epsilon_{\rm min}$ for the epochs with convex X-ray spectrum.  
Blue, green and red lines correspond to the observation IDs 0158961401, 0411780101 and 0411780201 respectively. 
Dashed horizontal lines indicate $1\sigma$ confidence level.}\label{fig:emin}
\end{figure}

In order to explore the minimum photon energy ($\epsilon_{\rm min}$) that could observationally be associated with SSC 
emission, we modify equation~(\ref{eq:doupow}) as
\begin{align}\label{eq:doupowemin}
	F_{\rm dp}(\epsilon)=F_{\rm 0}\left[ \left(\frac{\epsilon}{\epsilon_{\rm m}}\right)^{-\Gamma_{\rm syn}} + 
		\left(\frac{\epsilon}{\epsilon_{\rm m}}\right)^{\Gamma_{\rm com}}\,\Theta(\epsilon-\epsilon_{\rm min}) \right]\,
\end{align}
where $\Theta$ is a Heaviside function. Using equation~(\ref{eq:doupowemin}) as a local model in XSPEC, and fixing $F_{\rm 0}$, $\Gamma_{\rm syn}$ and $\Gamma_{\rm com}$ to their best fit values, we then determined an upper limit $\epsilon_{\rm min,c}$ such that $\epsilon_{\rm min} < \epsilon_{\rm min,c}$ did not modify the fit statistics. 
In Table \ref{tab:emin}, we provide the 1-$\sigma$ upper limit on the $\epsilon_{\rm min,c}$ and in Figure \ref{fig:emin}
we show the variation of $\chi_{\rm red}^2$ with $\epsilon_{\rm min}$.
\begin{table}
	\begin{center}
	\begin{tabular}{c c c}
      	\hline
      	Obs. ID & $\epsilon_{\rm min,c}$ & $\chi_{\rm red}^2$ \\ 	
       \hline
0158961401  & 0.607  & 1.064 \\ 
0411780101  & 0.619  & 1.040 \\
0411780201  & 0.654  & 1.128 \\
        \hline
      	\end{tabular}
	\end{center}
	\caption{Upper limit on the minimum photon energy of the Compton spectra ($\epsilon_{\rm min,c}$) inferred
	by using a double power-law fit (see text).}
   	\label{tab:emin}
\end{table}
These findings suggest that the IC power-law
can be extended down to at least $\approx 0.6$ keV. It is tempting to extend this further down as the \emph{XMM-Newton} EPIC instrument is sensitive from 0.15 keV. However, we focused only on 0.6-10 keV range, since there can be contamination due to low pulse height events at the low energy end of the PN camera.

\section{Minimum electron energy and Jet Energetics}\label{sec:jeteng}
Constraints on $\epsilon_{\rm min,c}$ can be used to gain insights into the minimum electron energy $\gamma_{\rm min}$ 
of the non-thermal electron distribution responsible for the broadband emission. However, for blazars this demands 
prior knowledge about the physical parameters of the source. Given a jet Doppler factor $\delta$ and a source magnetic 
field strength $B$, the characteristic mean energy change in IC/SSC scattering (Thomson regime), considering 
relativistic and cosmological effects, will be \citep{1986rpa..book.....R}
\begin{align}
  \epsilon_{\rm SSC}(\gamma) 
  = \left(\frac{\delta}{1+z}\right)\gamma^2 \epsilon_{\rm syn}^\prime\,, 
\end{align}
where $z$ is the source red shift and $\epsilon_{\rm syn}^\prime$ the (synchrotron) soft photon energy. The 
characteristic synchrotron photon energy corresponding to the electron Lorentz factor $\gamma$, as measured 
in the rest frame of the emitting region, is given by 
\begin{align}
	\epsilon^{\prime}_{\rm syn}(\gamma) = \gamma^2 \left(\frac{he B}{2\pi m_{\rm e} c}\right)\,. 
	\nonumber
\end{align}
Requiring the Compton branch to extend down to at least $\epsilon_{\rm min,c}$, i.e. $\epsilon_{\rm 
SSC}(\gamma_{\rm min}) \lesssim \epsilon_{\rm min,c}$, one can obtain an upper limit on 
$\gamma_{\rm min}$ as 
\begin{align}\label{eq:eminul}
	\gamma_{\rm min} \lesssim \left[ \frac{2\pi m_{\rm e} c (1+z)}{he\delta B}\epsilon_{\rm min,c}\right]^{1/4}\,.
\end{align}
While there will be synchrotron (and correspondingly, SSC) emission below the associated $\gamma_{\rm min}$, 
approximately rising as $F_{\nu} \propto \nu^{1/3}$, this will (given the SED shape of PKS 2155-304) not much alter
the above noted constraint. Similarly, a lower limit on $\gamma_{\rm min}$ might be obtained in our case by requiring
that IC scattering of synchrotron (SED) peak photons does not lead to a contribution in the valley energy range. 
This constraint can be expressed as
\begin{align}\label{eq:eminll}
	\epsilon_{\rm min,c} < \gamma_{\rm min}^2 \epsilon_{\rm syn,p}\;,
\end{align}
where $\epsilon_{\rm syn,p}$ is the observed synchrotron spectral peak energy. Hence, equations~(\ref{eq:eminul}) 
and (\ref{eq:eminll}), suggest that the low-energy cutoff $\gamma_{\rm min}$ of the radiating electron distribution 
satisfies
\begin{align}\label{eq:emin}
	\sqrt{\frac{\epsilon_{\rm min,c}}{\epsilon_{\rm syn,p}}} 
	\lesssim \gamma_{\rm min} 
	\lesssim \left[ \frac{2\pi m_{\rm e} c (1+z)}{he\delta B}\epsilon_{\rm min,c}\right]^{1/4}\,.
\end{align}
The above relation implies that with suitable knowledge of $\delta$, $B$ and $\epsilon_{\rm syn,p}$, we can 
constrain $\gamma_{\rm min}$ through $\epsilon_{\rm min,c}$ obtained from the convex X-ray spectrum. 
As we show in Sec.~\ref{sec:sedfit}, the heuristic arguments employed here to infer $\gamma_{\rm min}$ are 
supported by results based on full spectral modelling.

\subsection{Kinetic Jet Power}
If we consider a distribution of radiating electrons that is approximated to a power-law described by\footnote{Broadband 
spectral fitting for PKS~2155-304 demands a broken power-law electron distribution and the same is considered in 
\S\ref{sec:sedfit}. Since the total electron number density is governed by the differential number density at $\gamma_{\rm min}$, the consideration of a simple power-law is sufficient here.}
\begin{align}
	N(\gamma) d\gamma = K \left(\frac{\gamma}{\gamma_{\rm p}}\right)^{-p} d\gamma; \quad \gamma_{\rm min}< \gamma < \gamma_{\rm p}\; ,
\end{align}
then, given $p>1$ and $\gamma_{\rm p} \gg \gamma_{\rm min}$, the total electron number density is dominated by the electrons with energy $\gamma_{\rm min}$ and is given by
\begin{align}\label{eq:ntot}
	N_{\rm tot} \approx \int\limits_{\gamma_{\rm min}}^{\gamma_{\rm p}} N(\gamma) d\gamma 
           \approx \frac{K}{p-1} \left(\frac{\gamma_{\rm min}}{\gamma_{\rm p}}\right)^{-p+1}\,.
\end{align}
Under the heavy (e-p) jet approximation, the mass density of the jet is dominated by protons whose 
number density is assumed to be equal (charge-neutrality) to that of the non-thermal electrons 
($N_{\rm tot}$). The bulk kinetic power of the jet ($P_{\rm jet,heavy}$) is then typically dominated
by protons (for $\gamma_{\rm min} \ll \left(\frac{m_{\rm p}}{m_{\rm e}}\right)$), and if we 
assume them to be cold, the jet kinetic power will be 
\begin{align}\label{eq:pjeth1}
	P_{\rm jet,heavy} \approx \pi R^2 \Gamma^2 \beta_{\rm \Gamma} U_{\rm p} c\,,
\end{align}	
where $U_{\rm p} = N_{\rm tot} m_{\rm p} c^2$ is the proton energy density, $R$ is the size of the jet region, 
$\Gamma$ is the bulk flow Lorentz factor and $\beta_{\rm \Gamma} =(1-\Gamma^{-2})^{1/2}$. Using 
equations (\ref{eq:emin}), (\ref{eq:ntot}) and (\ref{eq:pjeth1}), $P_{\rm jet,heavy}$ is constrained as
\begin{align}
	\left[\frac{he}{2\pi m_{\rm e} c(1+z)}\frac{B}{\epsilon_{\rm min,c}}\right]^{\frac{p-1}{4}} \delta^{\frac{p+7}{4}} 
	\lesssim \zeta P_{\rm jet,heavy} 
	\lesssim  \delta^2 \left(\frac{\epsilon_{\rm syn,p}}{\epsilon_{\rm min,c}}\right)^{\frac{p-1}{2}}
\end{align}
where
\begin{align}
	\zeta = \left(\frac{p-1}{\pi m_{\rm p} c^2 R^2 K}\right) \gamma_{\rm p}^{1-p} \nonumber\,,
\end{align}
and we assumed $\Gamma \approx \delta$ and $\beta_{\rm \Gamma} \approx 1$. 

For a light jet, where the energy density is dominated by electrons and positrons, the 
kinetic power of the jet ($P_{\rm jet,light}$) becomes
\begin{align}
	P_{\rm jet,light} \approx \pi R^2 \Gamma^2 \beta_{\rm \Gamma} U_{\rm e} c \,,
\end{align}
where $U_{\rm e}$ is the leptonic energy density given by 
\begin{align}\label{eq:ue}
	U_{\rm e} \approx m_{\rm e} c^2 \int\limits_{\gamma_{\rm min}}^{\gamma_{\rm p}} \gamma\,N(\gamma) d\gamma 
	      \approx \frac{K \gamma_{\rm p} m_{\rm e} c^2}{p-2} \left(\frac{\gamma_{\rm min}}{\gamma_{\rm p}}\right)^{-p+2} \,,
\end{align}
under the approximations $\gamma_{\rm p} \gg \gamma_{\rm min}\; {\rm and}\; p>2$.
Hence,
\begin{align}\label{eq:phl}
	P_{\rm jet,light} 
	\approx \left(\frac{p-1}{p-2}\right)\left(\frac{m_{\rm e}}{m_{\rm p}}\right)\,\gamma_{\rm min}\, P_{\rm jet,heavy}\,. 
\end{align}
Using equation (\ref{eq:emin}), $P_{\rm jet,light}$ is constrained by 
\begin{align}
	\left[\frac{he}{2\pi m_{\rm e} c(1+z)}\frac{B}{\epsilon_{\rm min,c}}\right]^{\frac{p-2}{4}} 
	\delta^{\frac{p+6}{4}} \lesssim \xi P_{\rm jet,light} 
	\lesssim  \delta^2 \left(\frac{\epsilon_{\rm syn,p}}{\epsilon_{\rm min,c}}\right)^{\frac{p-2}{2}}\,,
\end{align}
where
\begin{align}
	\xi = \left(\frac{p-2}{\pi R^2 K m_{\rm e} c^2}\right) \gamma_{\rm p}^{1-p}\,.
\end{align}


\section{Broadband Spectral Modelling}\label{sec:sedfit}

As noted above, estimation of the minimum available energy $\gamma_{\rm min}$ and the jet power demands 
knowledge of the source parameters, $\delta$ and $B$. In order to obtain these information, we performed 
a broadband spectral fitting of the SED considering both synchrotron and SSC emission processes \citep{2018RAA....18...35S}. 
The emission from the source is imitated by assuming the emission region to be a spherical region of size $R$, 
permeated by a tangled magnetic field $B$ and populated by a broken power-law distribution of the form
\begin{align}
N(\gamma) = K \times \left\{
\begin{array}{ll}
	\left(\frac{\gamma}{\gamma_{\rm p}}\right)^{-p},&\mbox {~$\gamma_{\rm min}<\gamma<\gamma_{\rm p}$~} \\\\
	\left(\frac{\gamma}{\gamma_{\rm p}}\right)^{-q},&\mbox {~$\gamma_{\rm p}<\gamma<\gamma_{\rm max}$~} \\
\end{array}
\right.
\end{align}
 The Doppler factor $\delta$ determines the flux enhancement due to the relativistic motion of the emission 
region along the jet. The observed flux due to synchrotron and SSC emission processes, after accounting for 
the relativistic and cosmological effects will be \citep[e.g.,][]{1984v56p255,1995ApJ...446L..63D}
\begin{align}
	F_{\rm obs}(\epsilon)= \frac{\delta^3(1+z)}{d_{\rm L}^2} V \left[
		j_{\rm syn}\left(\frac{1+z}{\delta_{\rm D}}\epsilon\right) + j_{\rm SSC}\left(\frac{1+z}{\delta_{\rm D}}\epsilon\right)\right]
\end{align}
where, $j_{\rm syn}$ and $j_{\rm SSC}$ are the synchrotron and SSC emissivities measured in the frame of 
emission region \citep[e.g.,][]{2018RAA....18...35S,2008ApJ...686..181F}, $d_{\rm L}$ is the luminosity distance 
and $V$ is the volume of the emission region. For the SSC emissivity, we have considered the full Klein-Nishina 
cross section for the scattering process. 
The spectrum is then essentially governed by seven source parameters namely, $p$, $q$, $K$, $\gamma_{\rm p}$, 
$R$, $B$ and $\delta$. The limited information available from the optical and X-ray energies does not
allow us to constrain all these parameters and hence, we impose certain conditions to determine the
characteristic range of the source parameters. We assume equipartition between the emitting electron 
energy density and the magnetic field. This assures a minimum energy budget \citep{1959ApJ...129..849B} 
and using equation~(\ref{eq:ue}) we can express $B$ as
\begin{align}\label{eq:eqB}
	B \approx \sqrt {\frac{8 \pi \eta K \gamma_{\rm p} m_{\rm e} c^2}{(p-2)} 
	\left(\frac{\gamma_{\rm min}}{\gamma_{\rm p}}\right)^{-p+2}}\,,
\end{align}
where $\eta = 1$ corresponds to equipartition and $\gamma_{\rm min}$ can be constrained from equation 
(\ref{eq:emin}). The parameter $\gamma_{\rm p}$ determines the SED peak of the synchrotron spectral
component 
\begin{align}
    \epsilon_{\rm syn,p} &= \left(\frac{h\delta}{1+z}\right) \left(\frac{e B}{2 \pi m_{\rm e} c}\right)\gamma_{\rm p}^2
\end{align}
and $\epsilon_{\rm syn,p}$ can be obtained from the intersection of extrapolated optical/UV and X-ray spectra.
Finally, we constrain $R$ by assuming a variability timescale, $t_{\rm var}$, as 
\begin{align}
	R\approx \frac{c}{1+z} \delta t_{\rm var}
\end{align}

Along with these constraints, we also included archival $\gamma$-ray data of the source 
as obtained by \emph{Fermi}-LAT for the spectral fit. Since simultaneous $\gamma$-ray observations were not available, we 
chose observations in 2008 \citep{2009ApJ...696L.150A} as this being the closest one available with the 
X-ray observation epochs (2006-2007) considered in this work. 
In addition, we also included 2013 $\gamma$-ray observations \citep{2016ApJ...831..142M} during 
which the source was found to be in its lowest flux state.

\begin{figure*}
    \centering
    \begin{subfigure}[l]{0.45\linewidth}
        \includegraphics[scale=0.3,angle=-90]{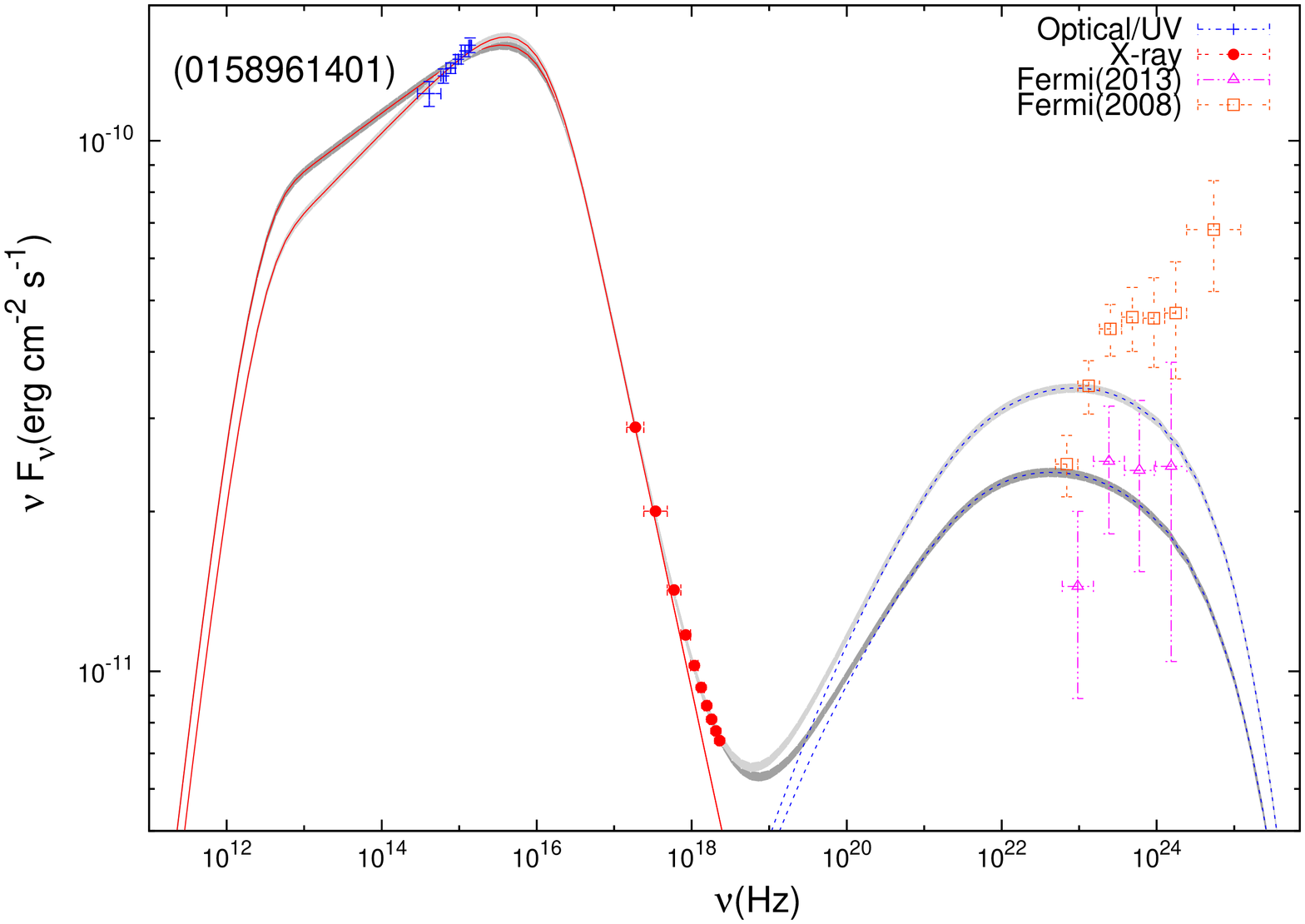}
    \end{subfigure}
        \vspace{4mm}
    \begin{subfigure}[l]{0.45\linewidth}
        \includegraphics[height=55mm,width=70mm,angle=0]{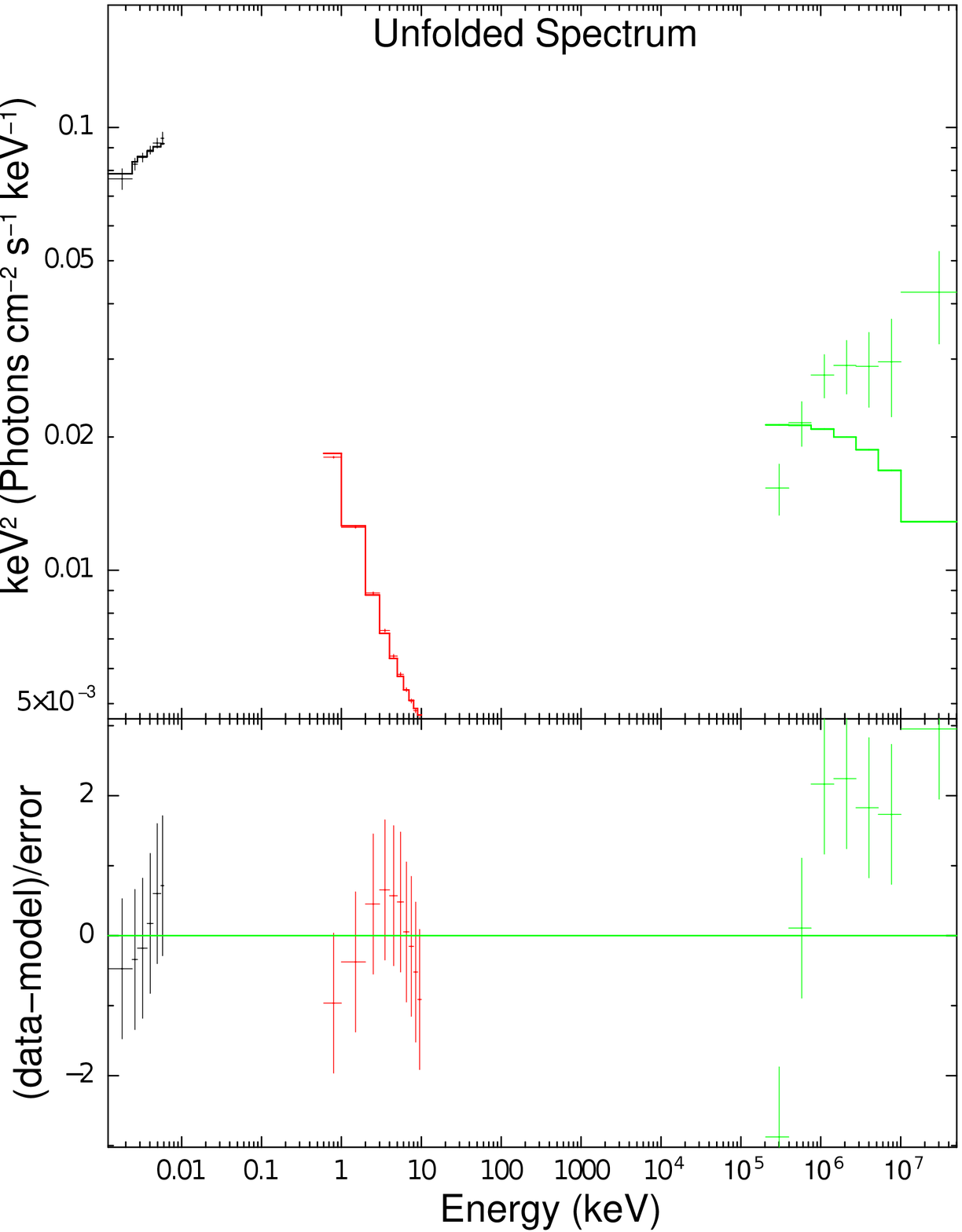}
    \end{subfigure}
    \centering
    \begin{subfigure}[]{0.45\linewidth}
        \includegraphics[scale=0.3,angle=-90]{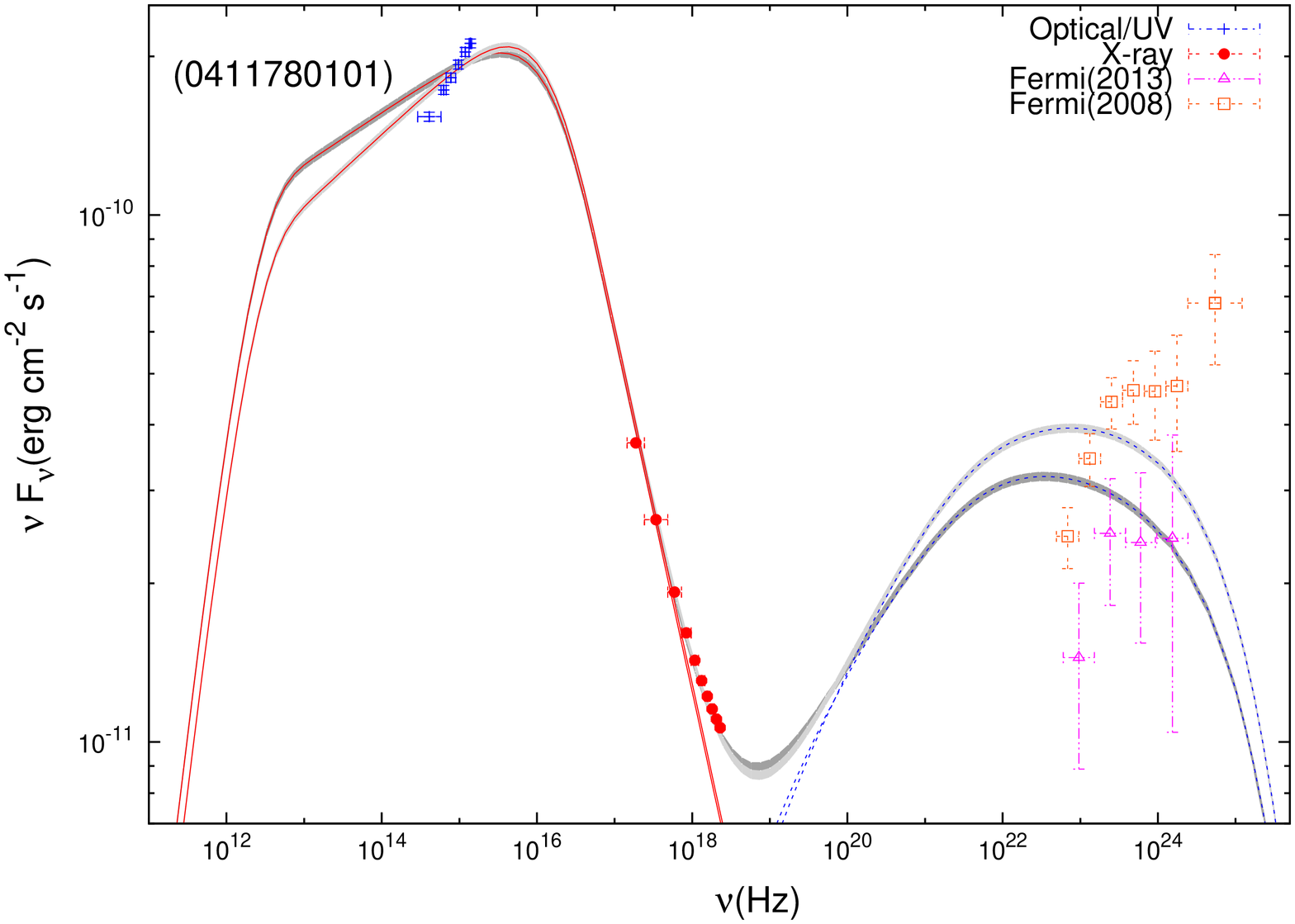}
    \end{subfigure}
        \vspace{4mm}
    \begin{subfigure}[]{0.45\linewidth}
        \includegraphics[height=55mm,width=70mm,angle=0]{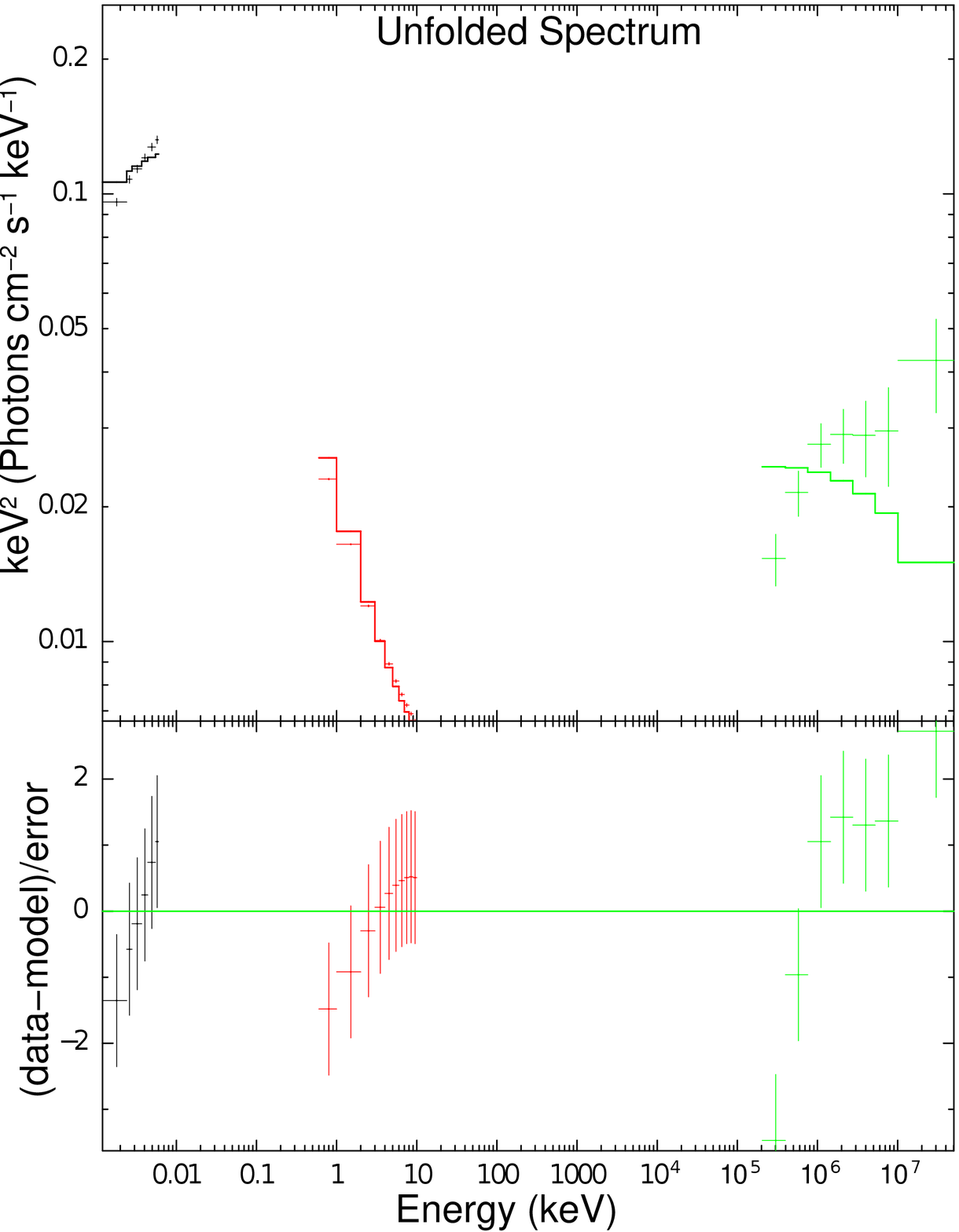}
    \end{subfigure}
    \centering
    \begin{subfigure}[]{0.45\linewidth}
        \includegraphics[scale=0.3,angle=-90]{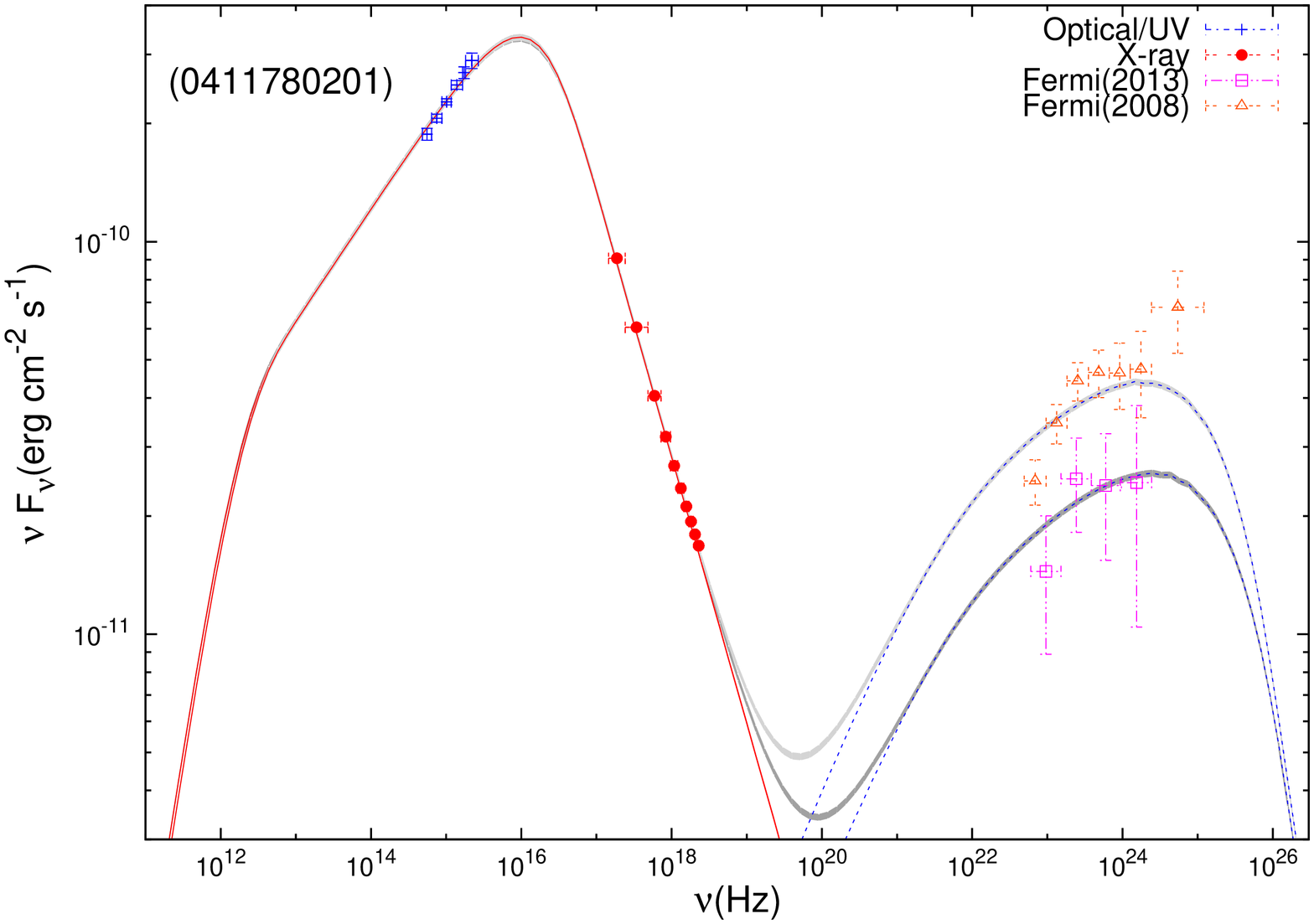}
    \end{subfigure}
    \begin{subfigure}[]{0.45\linewidth}
        \includegraphics[height=55mm,width=70mm,angle=0]{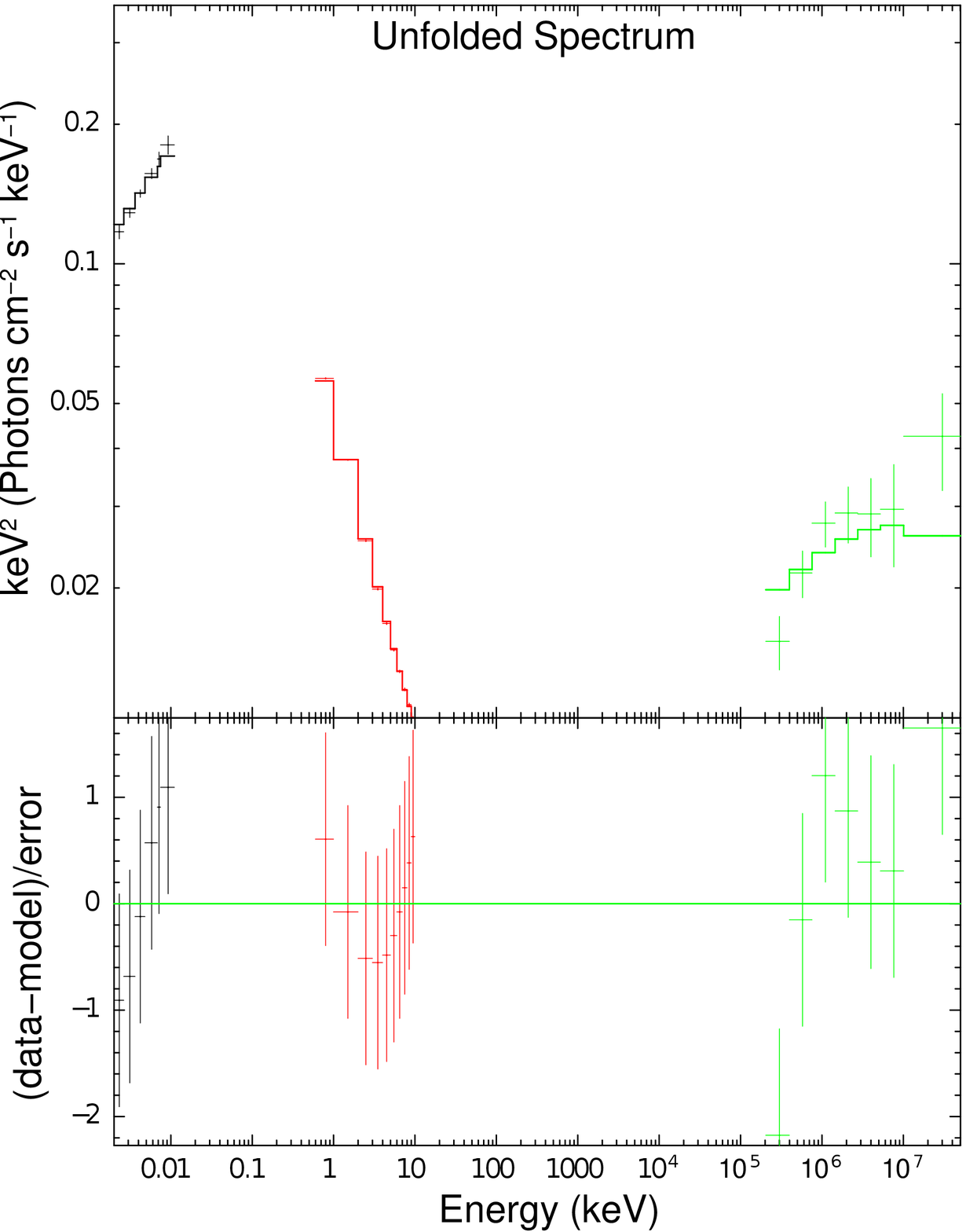}
    \end{subfigure}
    \caption{The best fit SEDs corresponding to three epochs with convex X-ray spectrum and the \emph{Fermi}-LAT reference 
observations of the source during 2008 and 2013. The solid line represents synchrotron and the dashed line represents SSC 
contribution. The thick gray band represents the combined emission with the width of the band representing the systematic 
error added to the model. The XSPEC spectral fit, along with residual are given in the right panel corresponding to the 2008 gamma-ray
observations.}\label{fig:bbsed1}
\end{figure*}

\begin{figure}
\includegraphics[scale=0.6,angle=270]{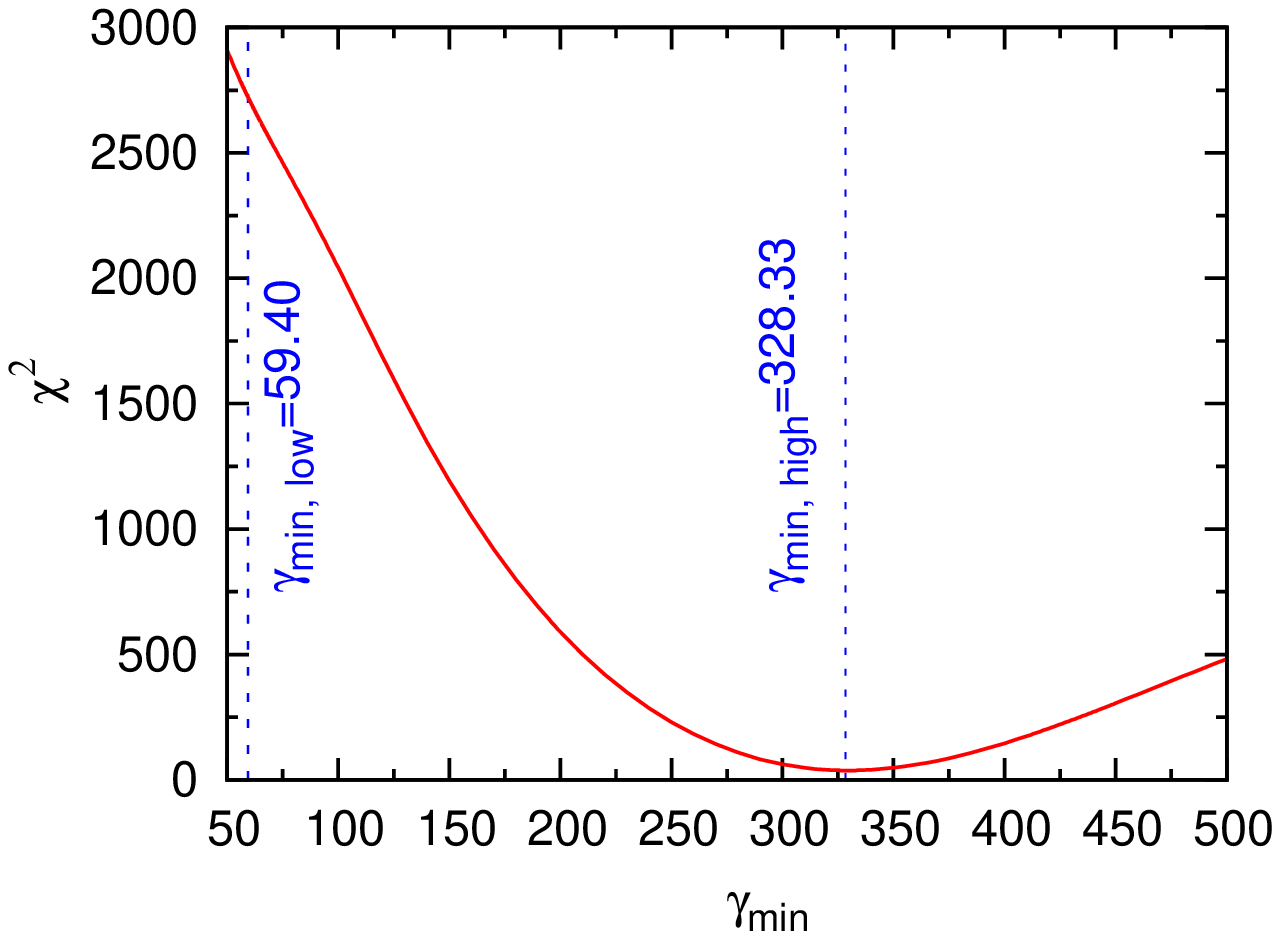}
\caption{Variation of $\chi^2$ of the spectral fit for the best fit parameters given in Table {\ref{tab:sed1}} 
(observation ID 0158961401) with change in $\gamma_{\rm min}$. The dashed vertical 
blue lines denote lower and upper limit on $\gamma_{\rm min}$ (cf. eq.~[\ref{eq:emin}]), respectively.\label{fig:eminfixvar}} 
\end{figure}

We developed a numerical code to generate the synchrotron and SSC emission spectra from the source 
parameters, along with the constraints discussed above. The computer routine is added as a local model in 
XSPEC and used to fit the optical/UV, X-ray and the archival $\gamma$-ray observations \citep{2018RAA....18...35S}.
The fitting procedure is performed as follows: We first fixed $t_{\rm var} = 1$ day, $\eta =1$ and 
$\epsilon_{\rm min,c}$ to values obtained from Table \ref{tab:emin}. Following this,
the best fit parameters were obtained by allowing the parameters $p$, $q$ and $\epsilon_{\rm syn,p}$ to vary 
within the confidence intervals obtained from the power-law/broken power-law/double power-law spectral fits to
optical/UV, X-ray and $\gamma$-ray spectra; whereas, $\delta$ and $B$ were set to vary freely. Next freezing 
the parameters to their best fit values except for $\delta$ and $B$, we obtained the confidence intervals on 
these two parameters. 
 In order to obtain error in the parameters, XSPEC demands $\chi_{\rm red}^2$ < 2. To achieve this we added 2 per 
cent additional error ({\it systematic} error in XSPEC) for the epochs corresponding to observation IDs 0158961401 and 
0411780101, and 7 per cent additional error for 0411780201.
The best fit source parameters corresponding to the two archival $\gamma$-ray observations are given in 
Table \ref{tab:sed1} and \ref{tab:sed2}. Using the best fit parameters, we obtained the range of $\gamma_{\rm min}$, 
$P_{\rm jet,heavy}$ and $P_{\rm jet,light}$ which are also provided in Table \ref{tab:sed1} and \ref{tab:sed2} 
(bottom). In the left panel of Figure \ref{fig:bbsed1}, we show the best fit broadband SED along with the 
observed fluxes and in right panel, the XSPEC fit results with the residuals corresponding to 2008 $\gamma$-ray 
observation. We note that the spectral fit is very sensitive to the choice of $\epsilon_{\rm min,c}$ or its equivalent 
$\gamma_{\rm min}$. In Figure \ref{fig:eminfixvar}, we show the variation of $\chi^2$ with $\gamma_{\rm min}$ while the rest 
of source parameters are fixed to their best-fit values for the spectral fit corresponding to the \emph{XMM-Newton} 
observation ID 0158961401. For the best fit parameters given in Table \ref{tab:sed1}, the minimum $\chi^2$ is 
achieved when $\gamma_{\rm min} \approx 330$. The required kinetic jet power for a light jet is of the order 
$3\times 10^{44}$ erg s$^{-1}$, and nearly an order of magnitude higher for a heavy jet. Assuming a black hole 
mass of $\sim5\times 10^{8} M_{\odot}$, this would roughly correspond to levels of $\sim0.1$ per cent and $\sim1$ 
per cent of the maximum possible jet power, respectively \citep[e.g.,][]{2018ApJ...852..112K}
%
\begin{table*}
	\begin{center}
	\begin{tabular}{l c c c}
	\hline
          Parameters &  (0158961401) & (0411780101) & (0411780201) \\ 
	\hline
$\delta$ & ${21.19}^{+0.71}_{-0.66}$ & ${23.94}^{+0.10}_{-0.10}$ & ${31.37}^{+1.31}_{-1.16}$ \\
B & ${0.238\pm0.013}$ & ${0.209}^{+0.002}_{-0.001}$ & ${0.114\pm0.008}$ \\
$p$ & 2.70 & 2.73 & 2.43 \\
$q$ & 4.36 & 4.36  & 4.36 \\
$\epsilon_{\rm min, c}$ & 0.607 & 0.619 & 0.654\\
$\epsilon_{\rm syn,p}$ ($10^{-2}$) & 3.31 & 3.64 & 5.02 \\

\hline
Properties &  &  &  \\
\hline
 $\gamma_{\rm min}$ & 59.40 - 328.33 & 67.27 - 330.45 & 60.61 - 364.66\\ 
 $P_{\rm jet, heavy}$ & 45.40 - 44.74  & 45.55 - 44.84  &  45.52 - 44.63  \\
 $P_{\rm jet, light}$ & 44.41 - 44.36  & 44.56 - 44.46 &  44.56 - 44.40 \\
\hline
\end{tabular}
	\end{center}
	\caption{The best fit source parameters and inferred quantities of PKS\,2155-304 for the three observation 
	IDs as obtained using a synchrotron and SSC emission model. The gamma-ray reference data are from 2008
	\emph{Fermi}-LAT observations \citep{2009ApJ...696L.150A}. $\delta$: Doppler factor, B: magnetic field (Gauss), 
	$p$ and $q$: low and high energy particle spectral indices, $\epsilon_{\rm min, c}$: observed minimum photon 
	energy of the Compton component (keV) (fixed to the values obtained from Table \ref{tab:emin}), $\epsilon_{\rm syn,p}$: 
	synchrotron peak energy (keV), $\gamma_{\rm min}$: minimum electron energy (in units of $m_{\rm e} c^2$), 
	$P_{\rm jet, heavy}$: jet kinetic power (log) estimated for an e-p jet (erg $\rm s^{-1}$), $P_{\rm jet, light}$: jet 
	kinetic power (log) estimated for a light jet (erg $\rm s^{-1}$). Fixed parameters are 
	$\eta=1$, $t_{\rm var} =1$ day and $\gamma_{\rm max} = 10^{7}$.}
	\label{tab:sed1}
\end{table*}

\begin{table*}
	\begin{center}
	\begin{tabular}{l c c c}
	\hline
Parameters &  (0158961401) & (0411780101) & (0411780201) \\ 
	\hline

$\delta$ & ${24.56\pm0.04}$ & ${26.10\pm0.08}$ & ${39.62\pm0.06}$ \\
B & ${0.1904\pm0.0005}$ & ${0.187\pm0.001}$ & ${0.0742\pm0.0002}$ \\
$p$ & 2.78 & 2.80 & 2.43 \\
$q$ & 4.36 & 4.36  & 4.36 \\
$\epsilon_{\rm min, c}$ & 0.607 & 0.619 & 0.654\\
$\epsilon_{\rm syn,p}$ ($10^{-2}$) & 3.70 & 3.81 & 5.03 \\

\hline
Properties &  &  &  \\
\hline
 $\gamma_{\rm min}$ & 77.70 - 334.44 & 92.17 - 332.36 & 60.55 - 382.64 \\ 
 $P_{\rm jet, heavy}$ & 45.44 - 44.82  & 45.50 -44.92  & 45.59 - 44.65 \\
 $P_{\rm jet, light}$ & 44.50 - 44.43  & 44.60 -44.52 &  44.61 - 44.44 \\
\hline

\end{tabular}
	\end{center}
	\caption{Same as in Table~\ref{tab:sed1}, but for gamma-ray reference data based on 2013 \emph{Fermi}-LAT 
	observations \citep{2016ApJ...831..142M}.}
	\label{tab:sed2}
\end{table*}
\newpage

\section{DISCUSSION AND CONCLUSION}\label{sec:discuss}

\begin{figure}
\includegraphics[scale=0.6,angle=270]{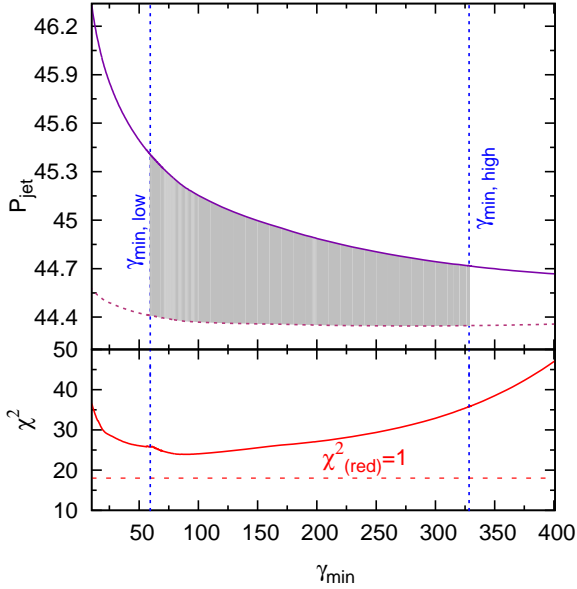}
\caption{Variation in jet power as a function of minimum electron energy $\gamma_{\rm min}$ (expressed in units of $m_{\rm e} c^2$) for the 
observation ID 0158961401. The solid purple line corresponds to $P_{\rm jet,heavy}$ and the dashed magenta line corresponds 
to $P_{\rm jet,light}$. The red solid line in the bottom panel is the best fit $\chi^2$ for different 
$\gamma_{\rm min}$ values and the dashed line denotes $\chi^{2}_{\rm red}=1$. The vertical blue dashed lines represent the limiting 
values of $\gamma_{\rm min}$. The gray shaded area is the allowed ranges of jet kinetic power.\label{fig:eminvar}} 
\end{figure}

Our analysis shows that the convex X-ray spectra of PKS\,2155-304 can be successfully interpreted as a 
combination of synchrotron and SSC spectral components. We demonstrate that this result along with the knowledge 
of source parameters can be effectively utilized to draw better constraints on the minimum energy $\gamma_{\rm min}$ 
of the emitting electron distribution and the power of the blazar jet. 
In general, the hard X-ray emission of PKS\,2155-304 is expected to reveal  a convex nature irrespective of flux state. 
 \emph{NuSTAR} observation of the source at 3-79 keV can thus effectively probe the valley regime and provide further 
constraints on $\gamma_{\rm min}$. In the current work, however, we wanted to highlight the occasional convex 
spectra that have been observed in the soft X-ray regime using \emph{XMM-Newton} observations. The constraints 
on $\gamma_{\rm min}$ are derived from these epochs of convex soft X-ray spectra. The source parameters were 
obtained through statistical fitting of the broadband SED using synchrotron and SSC emission processes. We observed 
that the spectral fit is very sensitive to the parameter $\gamma_{\rm min}$. To explore in detail how the jet power depends 
on $\gamma_{\rm min}$, we replaced the parameter $\epsilon_{\rm min,c}$ with $\gamma_{\rm min}$ in our numerical 
emission model and repeated the fitting procedure (as mentioned in \S \ref{sec:sedfit}). In Figure \ref{fig:eminvar}, we 
show the variations of $P_{\rm jet,heavy}$ (solid purple line) and  $P_{\rm jet,light}$ (dashed magenta line) with respect 
to $\gamma_{\rm min}$ along with the goodness-of-fit $\chi^2$ in the bottom (solid red line). The fitting was performed 
on the SED corresponding to the \emph{XMM-Newton} observation ID 0158961401 along with its 2008 $\gamma$-ray 
reference spectrum. The horizontal red dashed line corresponds to reduced $\chi^2 = 1$ and the vertical blue dashed 
lines corresponds to the limiting values of $\gamma_{\rm min}$ (Table  \ref{tab:sed1}). The dependence of $\chi^2$ over 
$\gamma_{\rm min}$ shown in Figure \ref{fig:eminfixvar} satisfies the condition obtained earlier in \S \ref{sec:jeteng} 
equation (\ref{eq:emin}). We found that $P_{\rm jet,heavy}$ varies more strongly with $\gamma_{\rm min}$ than 
$P_{\rm jet,light}$. Also, as $\gamma_{\rm min}$ approaches $\left(\frac{m_{\rm p}}{m_{\rm e}}\right)$, $P_{\rm jet,light} \approx 
P_{\rm jet,heavy}$. Further, considering the best fit values of $\delta$ and $B$, the allowed range of jet power can be 
constrained by the gray-shaded area in figure \ref{fig:eminvar}.

\begin{figure}
\includegraphics[scale=0.6,angle=270]{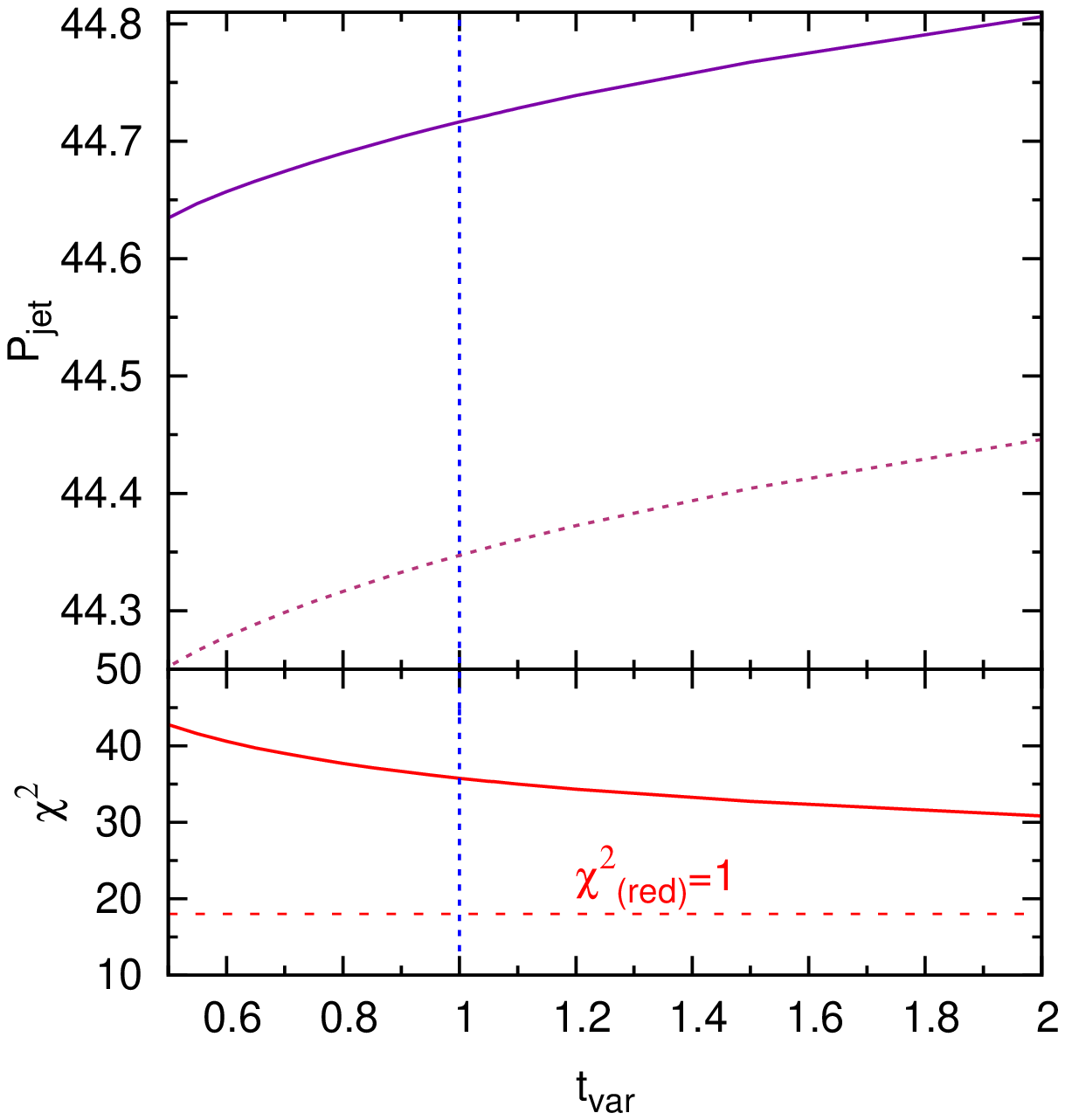}
\caption{Variation in jet power for a range of variability time-scale $t_{\rm var}$ (in days) corresponding to the observation 
ID 0158961401. The solid purple line corresponds to $P_{\rm jet,heavy}$ and the dashed magenta line corresponds to 
$P_{\rm jet,light}$. The red solid line in the bottom panel is the best fit $\chi^2$ for different $t_{\rm var}$ values and the 
dashed line denotes $\chi^{2}_{\rm red}=1$. The vertical blue dashed line represents $t_{\rm var} =1$ day.
\label{fig:emintvar}} 
\end{figure}

In general, the non-availability of simultaneous $\gamma$-ray data introduces some uncertainty on the fit 
parameters. On the other hand, Table \ref{tab:sed1} and \ref{tab:sed2} reveal that the estimates for
$\gamma_{\rm min}$ and the jet kinetic power do not vary much for the two different gamma-ray flux states of 
the source. The obtained limits on $\gamma_{\rm min}$ and jet powers may therefore be viewed as typical 
values for PKS\,2155-304. In principle, uncertainty in the size of the emission region $R$ could also impact
these estimates. To test this, we repeated the fitting by freezing $t_{\rm var}$ to a predefined value while 
setting $\eta =1$ and $\gamma_{\rm min}=328.33$. Figure \ref{fig:emintvar} shows the variation in 
$P_{\rm jet,light}$ and $P_{\rm jet,heavy}$ along with the $\chi^2$ (bottom panel). The variation in jet 
power is found to be within the same order while $t_{\rm var}$ changes from 0.1 to 2 days. This suggests 
that our inferred values represent the typical parameter space of the source.

The assumption of equipartition between the particle energy density and magnetic field relies on the principle 
that the source remains in its minimum energy state. However, there is no consensus that this condition should 
be satisfied rigorously. For instance, the chosen equipartition condition assumes that the radiating particle distribution 
largely determines the total electron density. However, protons may also carry significant energy (while still 
being radiatively inefficient) and for such conditions $\eta \ne 1$. Similarly, when the electrons undergo 
non-radiative losses (e.g., adiabatic expansion) along with synchrotron losses, the equipartition condition 
may not be satisfied \citep{2011ApJ...740...64L}. Besides this, broadband spectral fitting of blazar SED by 
synchrotron and IC emission also indicates significant deviation from equipartition 
\citep[e.g.,][]{2016MNRAS.456.2374T}. To study the dependence of jet power on the assumed equipartition, 
we also repeated fitting for different values of $\eta$ keeping $t_{\rm var} = 1$ and $\gamma_{\rm min}=328.33$. 
We found that the jet power varies by a factor of 10 as $\eta$ varies from 0.1 to 10. This large variation advocates 
$\eta$ to be an important parameter in deciding the energetics of the source.

It is widely accepted that the broadband non-thermal emission from blazars originates from relativistic jets. 
The formation and collimation of these jets, however, still remains an unresolved puzzle. Assumptions regarding 
this include jets that are solely powered by the accretion disk \citep{1982MNRAS.199..883B} or with additional 
energy derived from the spin of the black hole \citep{1977MNRAS.179..433B}. In both the cases, a crucial role is 
played by the magnetic field transporting power from the black hole/accretion disk to the jet 
\citep[e.g.,][]{2011MNRAS.418L..79T,2018ApJ...852..112K}. Since the magnetic field is thought to be supported 
by accretion, the jet power is supposed to be correlated to accretion power. Requiring that the power of heavy 
jets satisfy the linear regression \citep{2014Natur.515..376G}
\begin{align}
	{\rm log} P_{\rm jet,heavy} = 0.92\, {\rm log}\left(\dot{M}c^2\right)+4.09
\end{align}
would imply accretion rates up to $\dot{M} \simeq 0.02~M_{\odot}/$yr, or $\sim 2\times 10^{-3} \dot{M}_{\rm Edd}$
for a black hole of mass $5\times10^8 M_{\odot}$, which seems feasible.

Our SED modelling favours a minimum Lorentz factor for the non-thermal electron distribution of $\gamma_{\rm min}
\gtrsim 60$, with a strong preference for a value around $\gamma_{\rm min} \simeq 330$ (\S \ref{sec:sedfit}). 
Since radiative cooling is slow for the inferred parameters, it seems likely that $\gamma_{\rm min}$ has to be related 
to the underlying particle acceleration mechanism. In general, diffusive shock acceleration is commonly regarded 
as one of the most promising mechanisms \citep[e.g.,][]{2004MNRAS.349..336K,2007Ap&SS.309..119R,2009MNRAS.400..330G}. 
Shocks are known to convert the kinetic energy of the incoming flow to the thermal energy, represented by a relativistic Maxwellian particle distribution peaking at energy $\gamma_{\rm p} = 2\theta$, where $\theta \equiv kT$, along with a non-thermal, power-law type component above $\gamma_{\rm p}$. Here, $k$ is the Boltzmann constant and $T$ is the effective plasma temperature. For a pure electron-positron plasma this suggests $\gamma_{\rm min} \sim \Gamma_{\rm s}$ for the downstream particle distribution, where $\Gamma_{\rm s}$ is the relative Lorentz factor between the shocked and the unshocked fluid. For a pure electron-proton (e-p) plasma, on the other hand, $\gamma_{\rm min}$ depends on the degree of thermal coupling between electrons and protons. For $T_{\rm e} =\zeta T_{\rm p}$, where $T_{\rm e}$ and $T_{\rm p}$ are the electron and proton plasma temperatures, and $\zeta \leq 1$, one finds 
\begin{align}
	\gamma_{\rm min} = 2 \left(\frac{m_{\rm p}}{m_{\rm e}}\right) \,\Gamma_{\rm s} \frac{\zeta}{1+\zeta}\,.
\end{align}
In the case of inefficient thermal coupling, $\zeta \simeq \left(\frac{m_{\rm e}}{m_{\rm p}}\right)$ and hence, $\gamma_{\rm min} \sim 2 \Gamma_{\rm s}$. On the other hand, for strong thermal coupling $\zeta \simeq 1$ and $\gamma_{\rm min} \sim \left(\frac{m_{\rm p}}{m_{\rm e}}\right) \Gamma_{\rm s} $. For the jet in blazars, such as PKS\,2155-304, one typically has $\Gamma_{\rm s} \sim \Gamma$ ($\approx \delta$) or $\Gamma_{\rm s}\sim$ a few, for acceleration at external or internal shocks respectively. Our broadband spectral fitting assuming synchrotron and SSC emission mechanisms suggests $\delta \sim 20$ (Table \ref{tab:sed1}). Hence,
our constraint on $\gamma_{\rm min}$ would tend to disfavour a pure $e^+e^-$ plasma, while our best-fit value $\gamma_{\rm min}
\simeq 330$ would only be consistent with a pure e-p plasma composition in case of some reduced thermal 
coupling at mildly relativistic internal shocks. While PIC simulations of unmagnetized, collisionless relativistic electron-ion 
shocks \citep{2008ApJ...673L..39S,2013ApJ...771...54S,2020Galax...8...33V} indicate a thermal coupling that is 
somewhat stronger ($\zeta \sim 0.2-0.5$), variations in magnetization, shock speed and orientation may possibly 
explain the difference. If confirmed, this would tend to favour an e-p composition, at least for the X-ray emitting part 
of the jet.

Our results show that the convex X-ray spectra of blazars offer important insights into the source energetics as well 
as the matter content of jets. As indicated above, a crucial uncertainty in the present work relates to the equipartition 
parameter $\eta$. An appropriate value of $\eta$ could be obtained through a detailed study of the particle 
acceleration process. For instance, if the particles are energized through shock acceleration then a fraction of the
shock energy is spent in enhancing the magnetic field. A clear understanding about the energy budget involved 
demands detailed numerical simulation of relativistic MHD jet flow. This should be further supplemented with detailed 
spectral and temporal behaviour which can provide inputs/tests for the simulations.
One such information can be the simultaneous broadband SED extending up to GeV/TeV 
gamma-ray energies during lower flux states, supplemented with the temporal behavior of the source. Future upcoming 
ground-based Cherenkov Telescope Array observations will have the capability to gather information from the source at 
VHE energies, even at low flux states. This along with the inputs from other wavebands could play an important role in 
providing valuable information regarding the source energetics.

\begin{figure}
\includegraphics[scale=0.6,angle=270]{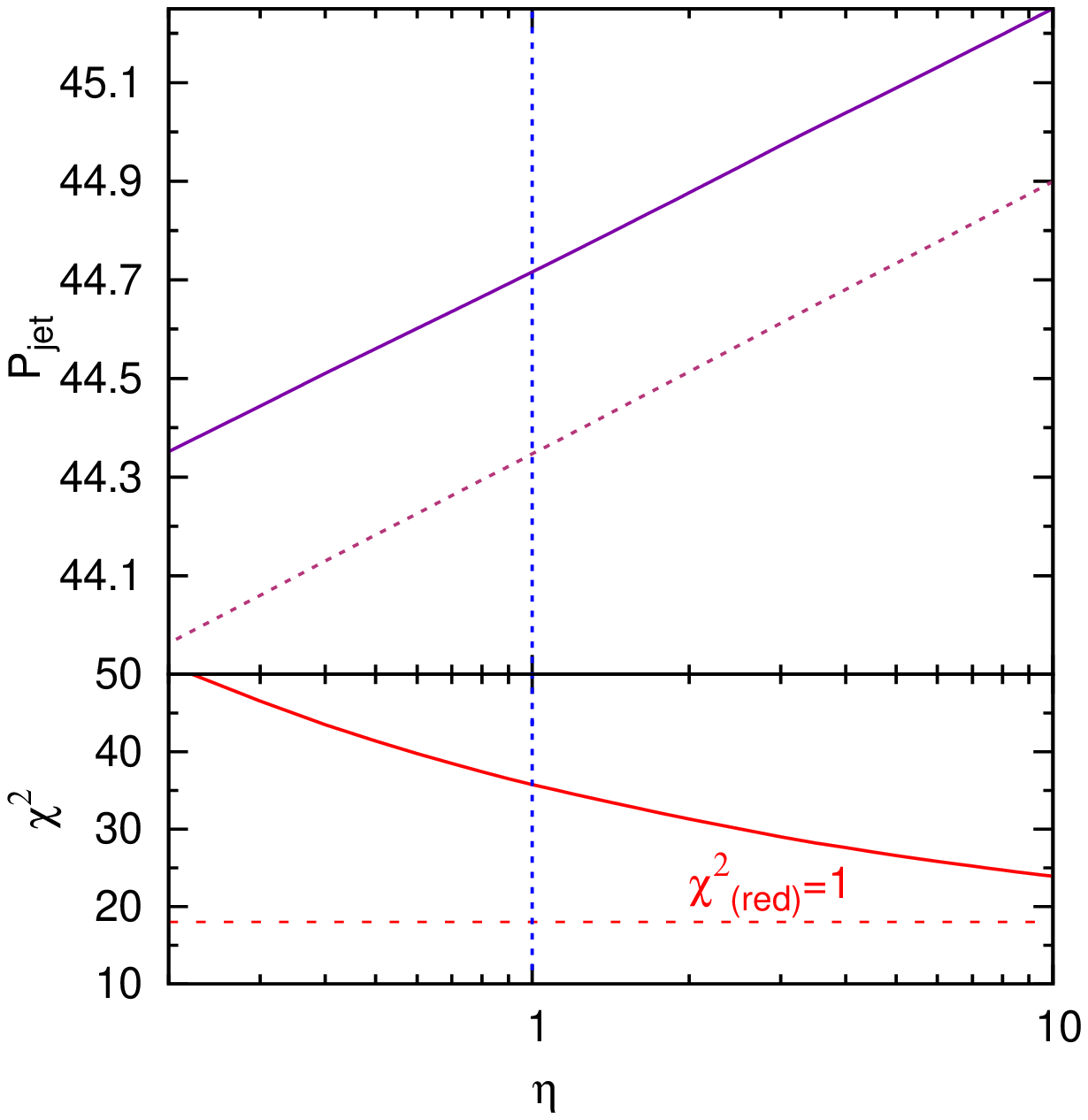}
\caption{Variation in jet power as a function of the equipartition parameter $\eta$ corresponding to the observation ID 0158961401. 
The solid purple line corresponds to $P_{\rm jet,heavy}$ and the dashed magenta line corresponds to $P_{\rm jet,light}$. The 
red solid line in the bottom panel is the best fit $\chi^2$ for different $\eta$ values and the dashed line denotes 
$\chi^{2}_{\rm red}=1$. The vertical blue dashed line represents $\eta=1$.
\label{fig:eminbeq}} 
\end{figure}




\section*{Acknowledgements}
Useful comments by the anonymous referee are gratefully acknowledged.
This research is based on observations obtained with the \emph{XMM-Newton} satellite, an ESA science mission 
with instruments and contributions directly funded by ESA member states and NASA. This work has made use of the 
NASA/IPAC Extra Galactic Database operated by Jet Propulsion Laboratory, California Institute of Technology and the 
High Energy Astrophysics Science Archive Research Center (HEASARC) provided by NASA's Goddard Space Flight 
Center. JKS thanks Jithesh V and Zahir Shah for useful discussions. JKS wishes to thank UGC-SAP and FIST 2 
(SR/FIST/PS1- 159/2010) (DST, Government of India) for the research facilities in the Department of Physics, 
University of Calicut. JKS is grateful to the University Grants Commission for the financial support through the RGNF 
scheme.  

\section*{Data Availability}
The data underlying this article are publicly available from the HEASARC\footnote{\url{https://heasarc.gsfc.nasa.gov/}} and the \emph{Fermi}-LAT\footnote{\url{https://fermi.gsfc.nasa.gov/ssc/data/access/}} archives.




\bibliographystyle{mnras}
\bibliography{references} 







\bsp	
\label{lastpage}
\end{document}